\documentclass[10pt, journal]{IEEEtran}
\usepackage{amsmath,amsfonts,amssymb,amsthm,bm,graphicx,multirow,subfigure,cite,stfloats}
\usepackage{array,enumerate,diagbox,makecell}
\usepackage{color}
\usepackage{changes}
\usepackage{algorithm, algorithmic,setspace}
\def\BibTex{{\rm B\kern-.05emf\sc i\kern-.025em b}\kern-.08emT kern-.1667em\lower.7ex\hbox{E}\kern-.125emX}

\begin{document}

\title{Multi-user Passive Beamforming in RIS-aided Communications and Experimental Validations}

\author{Zhibo Zhou, Haifan Yin,~\IEEEmembership{Senior Member,~IEEE}, Li Tan, Ruikun Zhang, Kai Wang, and Yingzhuang Liu
	\thanks{Z. Zhou, H. Yin, L. Tan, R. Zhang, K. Wang, Y. Liu are with the School of Electronic Information and Communications, Huazhong University of Science and Technology, Wuhan, China. E-mail: \{zbzhou, yin, ltan, zhangrk, kaiw, liuyz\}@hust.edu.cn.}
		\thanks{The corresponding author is Li Tan.}
		\thanks{This work was supported by the National Natural Science Foundation of China under Grants 62071191, 62071192 and 1214110.}
}
\maketitle

\begin{abstract}
Reconfigurable intelligent surface (RIS) is a promising technology for future wireless communications due to its capability of optimizing the propagation environments. Nevertheless, in  literature, there are few prototypes serving multiple users. In this paper,  we propose a whole flow of channel estimation and beamforming design for RIS, and set up an RIS-aided multi-user system for experimental validations. Specifically, we combine a channel sparsification step with generalized approximate message passing (GAMP) algorithm, and propose to generate the measurement matrix as Rademacher distribution to obtain the channel state information (CSI). To generate the reflection coefficients with the aim of maximizing the spectral efficiency, we propose a quadratic transform-based low-rank multi-user beamforming (QTLM) algorithm. Our proposed algorithms exploit the sparsity and low-rank properties of the channel, which has the advantages of light calculation and fast convergence. Based on the universal software radio peripheral devices, we built a complete testbed working at $5.8\text{ GHz}$ and implemented all the proposed algorithms to verify the possibility of RIS assisting multi-user systems. Experimental results show that the system has obtained an average spectral efficiency increase of $13.48\text{ bps/Hz}$, with respective received power gains of $26.6\text{ dB}$ and $17.5\text{ dB}$ for two users, compared with the case when RIS is powered-off. 
\end{abstract}

\section{Introduction}
\IEEEPARstart{R}{econfigurable} intelligent surface (RIS) is anticipated to be a potential key technology for future 6G mobile communication systems due to its great capability of manipulating the electromagnetic environment \cite{wu2019towards,di2020smart,pan2021reconfigurable}. By individually configuring the reflection coefficient of each element, RIS can efficiently execute a multitude of essential functions in communication, e.g., reflecting the incident radio signal towards a desired direction to facilitate additional beamforming, enhancing the rank of the channel to attain the full multiplexing gain \cite{rajatheva2020white}, suppressing co-channel interference \cite{liaskos2018new}, etc, while maintaining a low energy consumption and cost-effective hardware deployment. Furthermore, as the central frequency of wireless communication systems advances towards the mmWave/subTHz range, the RIS emerges as a potential attempt to play an increasingly pivotal role in the future. This potential stems from the pronounced power attenuation in free space exhibited by high-frequency electromagnetic waves, as well as the high penetration loss. The attractive benefits offered by RIS have spurred researches in various perspectives, such as element designs \cite{cui2014coding,zhang2018space,zhang2019breaking,dai2020reconfigurable,pei2021ris}, path loss modelings \cite{tang2020wireless,wang2021received,danufane2021path}, and transmission protocols \cite{lin2020adaptive,zheng2019intelligent,yang2020intelligent}. 

In practical terms, the challenge of RIS deployments lies mainly in how to configure it correctly to maximize its performance gains \cite{wu2019intelligent,guo2019weighted,ma2021joint}. 
Nevertheless, as far as our current understanding goes, only a limited number of prototypes have incorporated adaptive beamforming algorithms, and even fewer have ventured into the realm of more complex multi-user scenarios. The reasons for this phenomenon include overly desirable preconditions, high complexity, etc. In existing prototypes of RIS-aided wireless systems, there are four main methods to configure the reflection coefficients: 1) Beam-searching-based method, 2) Iterative algorithm, 3) Codeword design from location information, 4) Beamforming based on the estimated channel state information (CSI).

For a beam-searching-based scheme, the work in \cite{ren2022configuring} chooses to observe the conditional expectation of the received power to select the best configuration within a randomly generated reflection coefficient set, which can be extended to multi-user scenarios by substituting the performance indicator with the sum-rate of the whole system.
In the second category, the authors in \cite{pei2021ris} leverage extra feedback links to acquire a performance indicator, e.g., Reference Signal Received Power (RSRP). Through activating different groups of elements and observing the variations of the RSRP, the iterative algorithm facilitates a step-by-step enhancement in the received power for an individual user within real-world scenarios. This approach maintains a relatively high level of performance while exhibiting a low complexity. 
In the third category, the main theory of designing codeword is to guarantee that the signal reflected by each element adds up coherently at the receiver, which can be implemented with the help of steering vector to compensate the phase shifting of the multiple paths between the transmitter and the receiver. The authors in \cite{ouyang2023computer} attached cameras to the system to assist in obtaining location information and calculated the phase to be compensated for each element. The authors in \cite{xiong2023ris} proposed a method of Riemann manifold optimization to generate codewords for the far-field scenarios. 
In the last category, an adaptive beamforming algorithm based on CSI estimated from Matching Pursuit (MP) algorithm was proposed and applied into a single-user scenario to verify the capability of RIS in \cite{amri2021reconfigurable}.

Most early designs in RIS prototype avoided complex algorithms to verify the steering capability of RIS technology in a single-user case. Nevertheless, in order to characterize the capability of RIS serving multi-user wireless communication systems, the configuration of RIS requires a more strict method. Considering the need of applying in the real prototype, we divide the problem of configuring RIS into two separate sub-problems, i.e., channel estimation and multi-user beamforming. 

The channel estimation problem in an RIS-aided system is tough due to the passive nature of the elements and the high dimensionality of the entire surface. Early algorithm research works focused on simple signal processing techniques such as the Least Square/Linear Minimum Mean Square Error (LS/LMMSE) estimation method based on pilots. These traditional methods do not effectively solve the challenge of high dimensionality. To further reduce the dimensionality, one possible method is to exploit the structure properties of the channel vector\cite{wei2021channel}. Along this line, we introduce the angular domain channel model and formulate the channel estimation with Compressed Sensing (CS) terminology. Specifically, we propose to generate the measurement matrix as Rademacher distribution to match the 1-bit quantized RIS prototype. Moreover, we utilize Expectation Maximization-Generalized Approximate Message Passage (EM-GAMP) algorithm to obtain the cascaded CSI in RIS model. Based on the estimated CSI, the central mission of multi-user beamforming algorithms is to design the reflection coefficients by formulating multi-user beamforming as an optimization in a mathematical expression, to reach different goals according to different application scenarios. However, the main challenges of these algorithms when applied to practical prototypes includes the non-convexity introduced by the objective function and quantized constraints, and the high dimensionality. To address these two problems, one possible way is to solve the discrete optimization problem corresponding to the RIS model through heuristic algorithms \cite{zhi2022power, peng2021analysis}. In this paper, we introduce the iterative framework and exploit the low-rank property \cite{yin2013coordinated} in the channel model respectively. The proposed quadratic transform-based low-rank multi-user beamforming (QTLM) algorithm allows us to update the auxiliary variables of each sub-problem until the spectral efficiency of the whole system converges. Due to the non-decreasing property and a determined upper bound of the objective function, the algorithm is assured to converge at a local optimal point. The proposed method has the possibility of applying to prototypes and the advantages of light calculation and fast convergence.

The main contributions of this paper are summarized as follows:
\begin{itemize}
\item To the best of our knowledge, few works have been done on RIS-aided multi-user prototypes. We have designed and implemented an RIS-aided multi-user wireless communication system based on Universal Software Radio Peripheral (USRP) devices. Through a series of experiments, the results unequivocally reveal a significant augmentation in the spectral efficiency of the entire system after the configuration of the RIS. Furthermore, we conducted measurements on the radiation pattern of the codeword generated from our algorithms, and the findings indicate the potential application of the algorithm in a multi-user scenario and its ability to generate multiple beams.

\item  We have designed the whole workflow of the channel estimation and multi-user beamforming algorithms which can be employed in our RIS prototypes. Specifically, we introduce the angular domain channel model and formulate the channel estimation problem into CS terminology. In order to match our 1-bit quantized RIS prototype, we propose to generate the sensing matrix as Rademacher distribution and combine it with the generalized approximate message passing (GAMP) algorithm. Based on the estimated channel vector, we formulate the multi-user beamforming optimization problem and propose a quadratic transform-based low-rank multi-user beamforming (QTLM) algorithm to compute the reflection coefficients. In the process of the algorithms, we exploit the low-rank property in the model to reduce the complexity and accelerate the convergence. The proposed algorithm shows superior performance compared with existing schemes.

\item Since the algorithm we have developed requires an accurate noise power value throughout its iteration, the noise floor in radio frequency (RF) devices is not sufficient for characterizing it. To address this issue, we introduce a novel approach that involves compensating the noise power through the utilization of the Receive Modulation Error Ratio (RxMER). This correction factor considers the impairments present in RF chains, thereby enhancing the precision of the noise power estimation. By implementing this correction, the received power gain of a two-user system attains an average improvement of $10\text{ dB}$, underscoring the efficacy of this corrective technique.
\end{itemize}

The remainder of this paper is organized as follows: Sec. \ref{sec 2} introduces the system model. The channel estimation problem is formulated and the related algorithm to solve it is described in Sec. \ref{sec 3}. In Sec. \ref{sec 4}, we  formulate the passive beamforming optimization problem in multi-user system, and develop an iterative algorithm that could be applied to the prototype. Sec. \ref{sec 5} presents the simulation results comparing the proposed algorithm with other approaches, while Sec. \ref{sec 6} introduces the experimental results in our testbed. Sec. \ref{sec 7} concludes this paper.

\emph{Notations:} We use the boldface lower-case letter to denote a vector, the boldface upper-case letter a matrix. Let $({\bf X})^{\mathrm{T}}$, $({\bf X})^{\ast}$, and $({\bf X})^{\mathrm{H}}$ denote the transpose, conjugate, and conjugate transpose of a matrix $\bf X$ respectively. ${\bf X}_{i,j}$ denotes the $i,j$-th element in matrix $\bf X$. $\Vert {\bf x}\Vert_0$ and $\Vert { \bf x}\Vert_2 $ denote the $\ell_0$-norm and the $\ell_2$-norm of a vector $\bf x$ respectively. Denote the phase vector of a complex vector $\bf x$ by $\angle{\bf x}$. And $\text{diag}(\bm{\alpha})$ denotes a diagonal matrix with vector $\bm\alpha$ at the main diagonal. The Kronecker product of two matrices $\bf X$ and $\bf Y$ is denoted by ${\bf X}\otimes {\bf Y}$. For a complex number $x$, $\Re{(x)}$ denotes its real part, and $|x|$ denotes its absolute value. For a random variable $X \sim\mathcal{CN}(\mu,\sigma^2)$ represents that $X$ follows Circularly Symmetric Complex Gaussian distribution (CSCG) with expectation $\mu$ and variance $\sigma^2$.

\section{System Model}
\label{sec 2}
\begin{figure}[tbp]
\centering
\includegraphics[width=\linewidth]{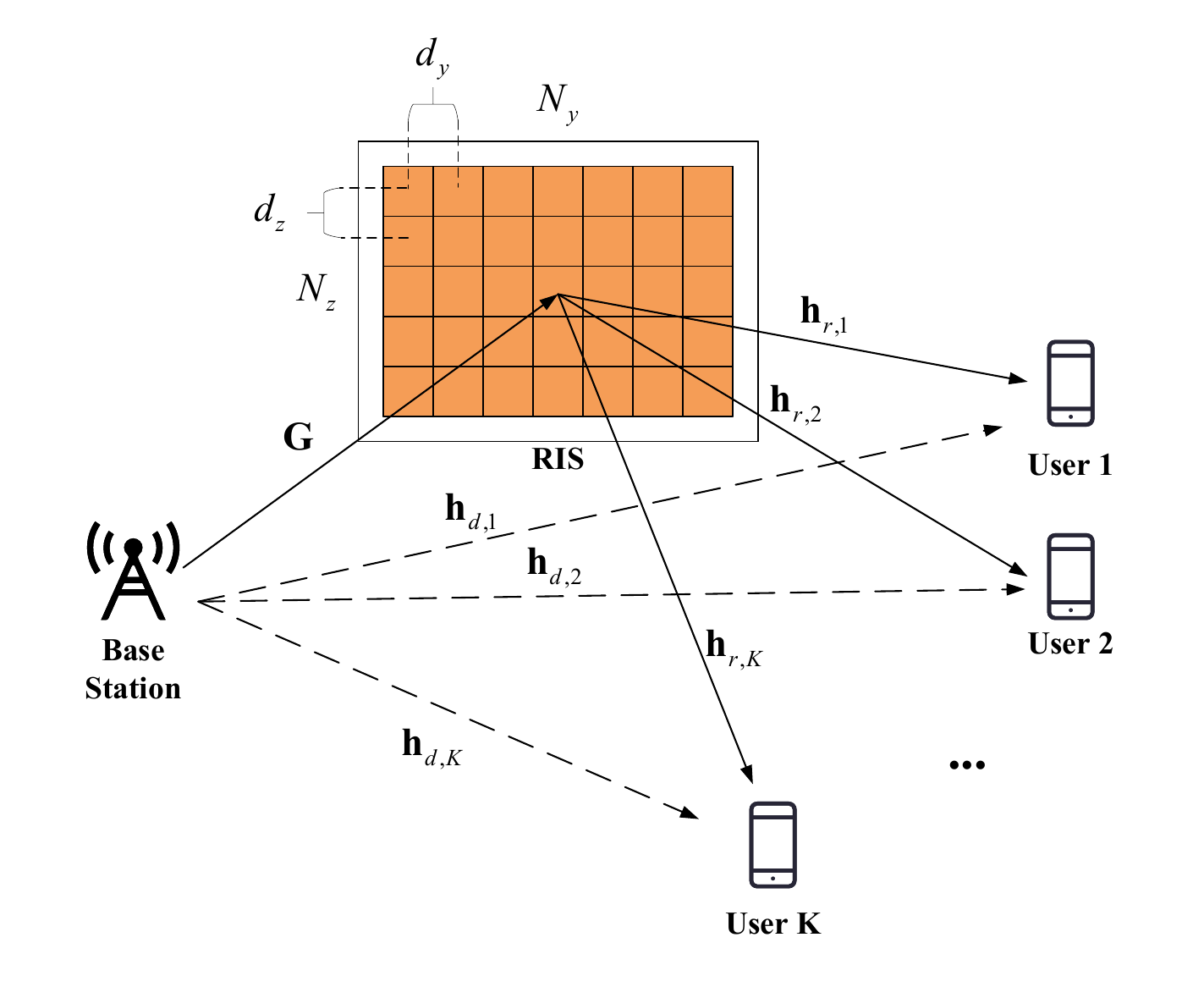}
\caption{An illustration of an RIS-aided multi-user wireless communication system.}
\label{systemFig}
\end{figure}

Consider an RIS-aided multi-user wireless communication system as shown in Fig. \ref{systemFig}, which consists of a base station (BS) with $M = M_yM_z$ antennas arranged in the form of uniform planar array (UPA), and $K$ users, equipped with a single antenna, where $M_y$ and $M_z$ are the numbers of antennas in the $y$-axis and the $z$-axis respectively. The elements in RIS form a UPA with $N_y$ columns and $N_z$ rows. The  number of elements in RIS is denoted by $N = N_yN_z$. The channels from the BS to the $k$-th user (also known as direct link), from the BS to the RIS, and from the RIS to the $k$-th user are denoted by ${\bf h}_{d,k}\in{\mathbb C}^{1\times M}$, ${\bf G}\in{\mathbb C}^{N\times M}$ and ${\bf h}_{r,k}\in{\mathbb C}^{1\times N}$ respectively, with $k=1,\cdots,K$. The reflection coefficients of the RIS elements are denoted by $\bm\theta=\begin{bmatrix}\theta_1&\theta_2&\cdots&\theta_N\end{bmatrix}^{\mathrm{T}}$. In engineering practice, the phase shifting of the element of the RIS is quantized to discrete bits. The phase shifting of a $\tau$-bit quantized RIS takes $2^\tau$values, i.e., $\theta_n=e^{j\varphi_n}$, where $\varphi_n\in\left \{0,\frac{1}{2^\tau}2\pi,\cdots,\frac{2^\tau-1}{2^\tau}2\pi\right \},n = 1,\cdots,N$. All the possible values of the discrete phase shifting form the set $\mathcal{F}_d$.

 In our model, the widely used Saleh-Valenzuela channel model \cite{saleh1987statistical,meijerink2014physical} is adopted to represent the direct channel between the BS and the $k$-th user as
\begin{equation}
    {\bf h}_{d,k} = \sum_{l}^{L_{d,k}}\beta_l^{d,k}{\bm\alpha}_M^{\mathrm T}(\vartheta_{l,\text{AoD}}^{\text{UE}}, \varphi_{l, \text{ZoD}}^{\text{UE}}),
\end{equation}
where $L_{d,k}$ denotes the number of paths between the BS and the $k$-th user, $\beta_l^{d,k}$ denotes the complex path gain of the $l$-th path, $\vartheta_{l,\text{AoD}}^{\text{UE}}$ and $\varphi_{l, \text{ZoD}}^{\text{UE}}$ denote the azimuth and elevation angles of departure of the $l$-th path from the UPA at the BS respectively. Furthermore, $\bm{\alpha}_M(\vartheta,\varphi)\in{\mathbb C}^{M\times 1}$ denotes the far-field steering vector for the UPA with $M = M_yM_z$ elements at the BS side. The array response of an $M=M_yM_z$ elements UPA can be represented as\cite{yin2020addressing}
\begin{equation}
{\bm\alpha}_M(\vartheta,\varphi)=\bm{\alpha}_y(\vartheta,\varphi)\otimes\bm{\alpha}_z(\varphi),
\end{equation}
where $\bm{\alpha}_y$ and $\bm{\alpha}_z$ denotes the array response vector of a uniform linear array along the $y$-axis and $z$-axis respectively:
\begin{equation}
\begin{aligned}
&\bm{\alpha}_y(\vartheta,\varphi)\\
&=\left [1,e^{-j2\pi\frac{d_y}{\lambda}\sin(\vartheta)\sin(\varphi)},\cdots,e^{-j2\pi\frac{d_y}{\lambda}\sin(\vartheta)\sin(\varphi)(M_y-1)}\right ]^{\mathrm{T}},
\end{aligned}
\end{equation}
\begin{equation}
\bm{\alpha}_z(\varphi)=
\left [1,e^{-j2\pi\frac{d_z}{\lambda}\cos(\varphi)},\cdots,e^{-j2\pi\frac{d_z}{\lambda}\cos(\varphi)(M_z-1)}\right ]^{\mathrm{T}},
\end{equation}
where $d_y$ and $d_z$ are $y$-axis and $z$-axis element spacing respectively, and $\lambda$ is the wave-length. 

Similarly, the channel $\bf G$  between the BS and the RIS can be represented as
\begin{equation}
    {\bf G}=\sum_{l}^{L_g}\beta_{l}^g\bm{\alpha}_N(\vartheta_{l,\text{AoA}}^{\text{RIS}},\varphi_{l,\text{ZoA}}^{\text{RIS}}){\bm\alpha}_M^{\mathrm T}(\vartheta_{l,\text{AoD}}^{\text{BS}}, \varphi_{l, \text{ZoD}}^{\text{BS}}),
\label{chan_g}
\end{equation}
where $L_g$ denotes the number of paths between the BS and the RIS, $\beta_{l}^g$ is the complex path gain of the $l$-th path, $\vartheta_{l,\text{AoA}}^{\text{RIS}}$ and $\varphi_{l,\text{ZoA}}^{\text{RIS}}$denote the azimuth and elevation angles  of arrival of the $l$-th path to the UPA at the BS respectively. $\vartheta_{l,\text{AoD}}^{\text{BS}}$ and $\varphi_{l, \text{ZoD}}^{\text{BS}}$ denote the azimuth and elevation angles of departure of the $l$-th path from the UPA at the BS respectively. ${\bm\alpha}_N(\vartheta,\varphi)\in{\mathbb C}^{N\times 1}$ denotes the array steering vector associated with the RIS of size $N=N_yN_z$. The element spacings in RIS model usually satisfy $d_y(d_z) \le \lambda/2$.

The channel ${\bf h}_{r,k}$ between the RIS and the $k$-th user is modeled as
\begin{equation}
{\bf h}_{r,k}=\sum_{l}^{L_k}\beta_{l}^k\bm{\alpha}_N^{\mathrm{T}}(\vartheta_{l,\text{AoD}}^{\text{RIS}},\varphi_{l,\text{ZoD}}^{\text{RIS}}),
\label{chan_h}
\end{equation}
where $L_k$ is the number of paths between the RIS and the $k$-th user, $\beta_{l}^k$ is the complex path gain of the $l$-th path, $\vartheta_{l,\text{AoD}}^{\text{RIS}}$ and $\varphi_{l,\text{ZoD}}^{\text{RIS}}$ denote the azimuth and elevation angles of departure from the UPA at the RIS. The distance boundary of near field and far field of the RIS is defined as \cite{tang2020wireless}
\begin{equation}
B = \frac{2N_yN_zd_yd_z}{\lambda}.
\label{bound}
\end{equation}

Denote the downlink transmitted signal to the $k$-th user by $s_k$, and normalize the power to $|s_k|=1$. Denote the beamforming vector adopted at the BS by ${\bf w}\in{\mathbb C}^{M\times 1}$, and the received signal of the $k$-th user is expressed as
\begin{equation}
\label{RISaid}
y_k=(\underbrace{{\bf h}_{d,k}{\bf w}s_k}_{\text{direct link}}+\underbrace{{\bf h}_{r,k}
{\text{diag}(\bm{\theta})}{\bf Gw}s_k}_{\text{RIS-aided link}})+n_k,
\end{equation}
where $n_k\sim{\mathcal{CN}(0,\sigma^2)}$. Note that in many commercial 5G macro base stations, hybrid beamforming schemes are employed with 8 fixed analog beams, each corresponding to a different angular direction. In the actual deployment scenarios of RIS, since the relative positions of the BS and the RIS generally do not change, we consider the analog beam that best fits the BS-RIS channel among these eight beams, and denote the selected beamforming vector at the BS by $\bf w$. This approach also sidesteps the tedious task of base station precoding design, significantly easing the complexity of implementing RIS in real-world systems. Let ${\bf h}_{d,k}{\bf w} = \hbar_{d,k}\in{\mathbb C}$ and ${\bf Gw} = {\bm \hbar}_g\in{\mathbb C}^{N\times 1}$ be the fixed channel part in the transmission model and let ${\bf h}_k=\text{diag}({\bf h}_{r,k}){\bm \hbar}_g\in{\mathbb C}^{N\times 1}$, then the received signal is
\begin{align}
y_k&=(\hbar_{d,k}+\bm{\theta}^{\mathrm{T}}\text{diag}({\bf h}_{r,k}){\bm \hbar}_g)s_k+n_k\\
\label{sig_model} &=(\hbar_{d,k}+\bm{\theta}^{\mathrm T}{\bf h}_k)s_k+n_k.
\end{align}
In this equation, the RIS-aided link in (\ref{sig_model}) can be divided into two parts, the first one is a vector of reflection coefficients, and the second only contains the fading channel, which is known as cascaded channel as well. Most studies assume that the direct link $\hbar_{d,k}$ is blocked by obstacles and ignores its effect on the received signal. However, our experimental results show that in the real environments this link could be small in power, compared with the RIS-aided link, yet  has an impact on how to configure the reflection coefficients of RIS. As a result, this part of the information is needed in our experiment.
\section{CS-based Channel Estimation Method}
\label{sec 3}

Most passive beamforming design research assumes perfect CSI and designs the beamforming algorithm based on this. In practice, however, the estimation of the cascaded channel is one of the major challenges of RIS. In our experiments, both the direct link channel and the cascaded channel need to be estimated.

The direct link channel estimate problem can be solved by utilizing conventional LS methods once we assume that RIS absorbs incident electromagnetic waves. In the  experiments described in Sec. \ref{sec 5}, the estimation of direct link channel will be conducted in a dedicated time slot. The main difficulty lies in the estimation of the cascaded channel due to the high dimension of the RIS board and the passive properties of the RIS element. To reduce the impact of pilot overhead on system throughput, we introduce an angular domain channel model. For the channel model in (\ref{sig_model}), the virtual angular expression is ${\bf h}_k^{\text{a}} = {\bf D}_N{\bf h}_k$, where ${\bf D}_N = {\bf D}_{N_y}\otimes{\bf D}_{N_z}$ is a Kronecker product of two discrete Fourier transform (DFT) matrices with dimensions $N_y$ and $N_z$. By taking this transform, only a few coefficients in ${\bf h}_k^{\text{a}}$ have relatively high magnitude, and other coefficients are close to zero, i.e., the angular-domain vector ${\bf h}_k^{\text{a}}$ is sparse, thus converting the original problem into estimating a sparse vector and significantly reducing the dimension of the signal to be estimated. The detailed algorithm is summarized below.

Assume that we estimate channel vector in $P$ time slots, with $P\ll N$. 
Denote the symbols transmitted in $P$ time slots to the $k$-th user by ${\bf s}=\begin{bmatrix}{}
s_1&s_2&\cdots&s_P
\end{bmatrix}^{\mathrm{T}}$, the reflection coefficients of RIS in the $p$-th slot are represented by ${\bm \theta}_p,p=1,\cdots,P$. All reflection coefficients vectors in $P$ slots form a matrix $\bm{\Theta}=\begin{bmatrix}
\bm{\theta}_1&\bm{\theta}_2&\cdots&\bm{\theta}_P
\end{bmatrix}\in{\mathbb {C}}^{N\times P}$. For the $k$-th user, the received signals in $P$ slots after removing the impact of the direct link are represented as 
\begin{equation}
{\bf y}_k=\begin{bmatrix}
\bm{\theta}_1^{\mathrm{T}}{\bf h}_{k}s_1+n_{k,1}\\
\bm{\theta}_2^{\mathrm{T}}{\bf h}_{k}s_2+n_{k,2}\\
\vdots\\
\bm{\theta}_P^{\mathrm{T}}{\bf h}_{k}s_P+n_{k,P}
\end{bmatrix}
=\begin{bmatrix}
  {\bm\theta}_1^{\mathrm T}{\bf h}_ks_1\\  
  {\bm\theta}_2^{\mathrm T}{\bf h}_ks_2\\ 
  \vdots\\
  {\bm\theta}_P^{\mathrm T}{\bf h}_ks_P\\ 
\end{bmatrix}+{\bf n}_k,
\end{equation}
where ${\bf n}_k$ is a vector of additive white Gaussian noise. Denote ${\bf M}={\bm{\Theta}}^{\mathrm{T}}{\bf D}_{N}^{\mathrm{H}}$ and we have
\begin{equation}
  {\bf y}_k={\bf M}{\bf h}_{k}^{\mathrm{a}}+{\bf n}_k.
  \label{compressed_sensing}
\end{equation}
The channel estimation problem is to reconstruct the channel vector ${\bf h}_k^{\mathrm{a}}$ from the known matrix ${\bf M}$ and the received signal vector ${\bf y}_k$. With the sparse structure of ${\bf h}_k^{\mathrm{a}}$, we can utilize CS framework to solve it. Formulate as an optimization problem:
\begin{equation}
\begin{aligned}
&\min_{{\bf h}_k^{\mathrm a}}\quad \Vert{\bf h}_k^{\mathrm{a}} \Vert_0&\\
& {\mathrm{s.t.}}\quad \Vert {\bf y}_k-{\bf Mh}_k^{\mathrm a}\Vert_2^2&\le \epsilon,
\end{aligned}
\label{CS_constr}
\end{equation}
where $\epsilon$ is the tolerance upper bound related to noise power. In CS terminology, ${\bf M}$ is the $P\times N$ sensing matrix, ${\bf y}_k$ is the $P\times 1$ measurement vector\footnote{Note that $P$ denotes the number of time slots used for channel estimation. However, in CS terminology, the definition of $P$ commonly refers to the number of samples. Thus, in the later description of experimental results, we have adopted similar terminology in CS.}.
The sensing matrix should satisfy Restricted Isometry Property (RIP) to recover sparse signal \cite{candes2006robust}, which guarantees that the columns of $\bf M$ are nearly orthonormal. In particular, it is shown that some random matrices such as Gaussian matrix, Bernoulli matrix, etc, satisfy the RIP with exponentially high probability\cite{tropp2007signal}. In our experimental model, since the RIS is quantified to one bit, the possible values of reflection coefficient (${\bm\Theta}_{i,j}\in\{1,-1\}$) match a Rademacher distribution \footnote{It has been shown in \cite{10.5555/2526243} that Rademacher random matrix belongs to the category of the sub-Gaussian matrices which satisfy RIP with high probability.}. Hence we propose to randomly generate $\bm \Theta$ as a Rademacher matrix. The probability mass function of ${\bm\Theta}_{i,j}$ is
\vspace{-1em}
\begin{equation}
f({\bm\Theta}_{i,j}) = \left \{\begin{aligned}
&\frac{1}{2} &\text{if }{\bm\Theta}_{i,j} = 1,\\
&\frac{1}{2} &\text{if }{\bm\Theta}_{i,j}=-1,\\
&0  &\text{otherwise.} 
\end{aligned} \right.
\end{equation}

Compared with the randomized distribution with continuous values, the Rademacher matrix retains less information after projecting into low-dimensional space since the available values of this distribution is limited in $\{1,-1\}$, resulting in performance loss in signal recovery algorithms. As a result, this loss stems unavoidably from the quantization limitation of RIS hardware system.

Mathematically, there are many methods to solve CS problems, such as Orthogonal Matching Pursuit (OMP), Basis Pursuit (BP), etc. Different algorithms will converge to different results (maybe local optimal) and have different complexity. In practice, we adopt the GAMP algorithm in the channel estimation, since it has the advantages of lightweight calculation and fast convergence. However, it also requires a prior information of the sparsity rate of the channel vector, which is unknown in real world. Therefore, we choose to combine it with the Expectation Maximization (EM) method to estimate the relevant parameters \cite{6556987}. 

\section{Multi-user Beamforming}
\label{sec 4}
\subsection{Problem Formulation}
Due to channel reciprocity in RIS-aided system in time division duplexing (TDD) mode \cite{pei2021ris}, the downlink channel can be obtained by the uplink channel. We will therefore not distinguish between uplink and downlink in our discussion. Note that due to the limitations of the hardware platform, we consider the setting of a single-antenna BS broadcasting signals to multiple single-antenna UEs. Based on the transmission model in (\ref{sig_model}), the received Signal-to-Noise Ratio (SNR) of the $k$-th user is computed by
\begin{equation}
\label{gamma}
\gamma_k=\frac{|\hbar_{d,k}+\bm{\theta}^{\mathrm{T}}{\bf h}_{k}|^2}{\sigma^2}.
\end{equation}
Then the spectral efficiency of the whole broadcast communication system is defined as
\begin{equation}
\label{sum_rate}
f({\bm{\theta}})=\sum_{k=1}^K\log_2(1+\gamma_k).
\end{equation}
Our objective is to configure the reflection coefficients of RIS elements $\bm{\theta}$ to maximize the spectral efficiency under discrete constraints, namely passive beamforming design. Formulate an optimization problem:
\begin{equation}
(\text{P1})\quad\begin{aligned}
&\max_{\bm{\theta}} &f_1({\bm{\theta}})&=\sum_{k=1}^K\log_2(1+\gamma_k)\\
&\mathrm{s.t.} &\theta_n&\in{\mathcal{F}}_d\forall n=1,\cdots,N.
\end{aligned}
\end{equation} 
For such an optimization problem, both the objective function and the constraints are non-convex. We propose an iterative algorithm based on Lagrangian dual transform and quadratic transform, reduce the original problem into convex one and derive the closed-form solution for each sub-problem, which highly reduces the complexity. The details of the algorithm are shown in the next subsection.

\subsection{Proposed Quadratic Transform-based Low-rank Multi-user Beamforming Algorithm}

To tackle logarithm in the objective function of (P1), we apply the Lagrangian dual transform \cite{shen2018fractional}, the problem (P1) can be equivalently written as

\begin{equation}
\text{(P1a)}\quad\begin{aligned}
&\max_{\bm{\theta},\bm{\alpha}} \quad f_{1a}({\bm{\theta}},{\bm{\alpha}})\\
&\begin{array}{r@{\quad}r@{}l@{\quad}l}
\text{s.t.} &\theta_n&\in \mathcal{F}_d \quad \forall n=1,\cdots,N,
\end{array}
\end{aligned}
\end{equation}
where $\bm{\alpha}=\begin{bmatrix}
\alpha_1&\alpha_2&\cdots&\alpha_K
\end{bmatrix}^{\mathrm{T}}\in{\mathbb{R}}^K$ is the auxiliary variables for the received SNR $\gamma_k$, and a new objective function is defined:
\begin{equation}
f_{1a}({\bm{\theta,\alpha}})=\sum_{k=1}^K\log_2(1+\alpha_k)-\sum_{k=1}^K\alpha_k+\sum_{k=1}^K\frac{(1+\alpha_k)\gamma_k}{1+\gamma_k}.
\label{f1a}
\end{equation}

In (P1a), when $\bm{\theta}$ is fixed, the optimization problem for $\bm \alpha$ is convex, thus by letting ${\partial f_{1a}(\bm{\theta,\alpha})}/{\partial \alpha_k}= 0$, the optimal $\alpha_k$ is obtained:
\begin{equation}
\label{alpha}
\alpha_k^{\text{opt}} = \gamma_k.
\end{equation}
Then for a fixed $\alpha_k$, define $\tilde{\alpha}_k=1+\alpha_k$, the variables of the optimization problem, i.e., $\gamma_k, k = 1, \cdots,K$, exist only in the third term in (\ref{f1a}). The optimization problem is reduced to
\begin{equation}
\label{OPT2}
\text{(P2)}\quad
\begin{aligned}
&\max_{\bm{\theta}} \quad f_{2}({\bm{\theta}})\\
&\begin{array}{r@{\quad}r@{}l@{\quad}l}
\text{s.t.} &\theta_n&\in {\mathcal F}_d \quad \forall n=1,\cdots,N,
\end{array}
\end{aligned}
\end{equation}
where
\begin{equation}
f_2({\bm{\theta}}) = \sum_{k=1}^K\frac{\tilde{\alpha}_k\gamma_k}{1+\gamma_k}.
\end{equation}

The problem (P2) is the sum of multiple-ratio fractional programming problem and non-convex for $\bm \theta$. Utilize the quadratic transform \cite{shen2018fractional} and substitute (\ref{gamma}) into (\ref{OPT2}), (P2) is equivalent to
\begin{equation}
\text{(P2a)}\quad
\begin{aligned}
&\max_{\bm{\theta},\bm{\varepsilon}} \quad f_{2a}({\bm{\theta}},\bm{\varepsilon})&=\sum_{k=1}^K2\sqrt{\tilde{\alpha}_k}\Re\left \{\varepsilon_k^\ast(\hbar_{d,k}+{\bm{\theta}^{\mathrm{T}}}{\bf h}_k)\right \}\\
&\quad&-\sum_{k=1}^K|\varepsilon_k|^2(\sigma^2+|\hbar_{d,k}+\bm{\theta}^{\mathrm{T}}{\bf h}_k|^2)\\
&\begin{array}{r@{\quad}r@{}l@{\quad}l}
\text{s.t.} &\theta_n&\in {\mathcal F}_d,
\end{array}
\end{aligned}
\end{equation}
where $\bm \varepsilon=\begin{bmatrix}
\varepsilon_1&\cdots&\varepsilon_K
\end{bmatrix}^{\mathrm{T}}\in{\mathbb C}^K$ is the auxiliary variable. Similarly, (P2a) for $\bm\varepsilon$ is convex, and by letting ${\partial f_{2a}(\bm{\theta,\varepsilon})}/{\partial\varepsilon_k}=0$, we derive the optimal value as
\begin{equation}
\label{varepsilon_opt}
\varepsilon_k^{\mathrm{opt}}=\frac{\sqrt{\tilde{\alpha}_k}\left(\hbar_{d,k}+{\bm{\theta}^{\mathrm{T}}{\bf h}_k}\right )}{\sigma^2+|\hbar_{d,k}+{\bm{\theta}^{\mathrm{T}}{\bf h}_k}|^2}.
\end{equation}
Then substitute (\ref{varepsilon_opt}) into $f_{2a}({\bm \theta},{\bm \varepsilon})$, we obtain
\begin{equation}
f_{2b}({\bm{\theta}}) = -{\bm\theta}^{\mathrm{T}}{\bf U}{\bm{\theta}^\ast}+2\Re(\bm{\theta}^{\mathrm T}{\bf v})+C,
\end{equation}
where
\begin{align}
{\bf U} &= \sum_{k=1}^K|\varepsilon_k|^2{\bf h}_k{\bf h}_k^{\mathrm H}, \\
{\bf v} &= \sum_{k=1}^K\left (\sqrt{\tilde{\alpha}_k}\varepsilon_k^\ast{\bf h}_k-|\varepsilon_k|^2\hbar_{d,k}{\bf h}_k\right ), \\
C &= \sum_{k=1}^K\left (\sqrt{\tilde{\alpha}_k}\Re(\varepsilon_k^\ast \hbar_{d,k})-|\varepsilon_k|^2\sigma^2-|\varepsilon_k|^2|\hbar_{d,k}|^2\right ).
\end{align}
The optimization problem becomes
\begin{equation}
\text{(P3)}\quad
\begin{aligned}
&\max_{\bm{\theta}}& f_{2b}&(\bm{\theta})\\
&\mathrm{s.t.}& \theta_n&\in{\mathcal F}_d.
\end{aligned}
\end{equation}
Since for an arbitrary non-zero vector ${\bf x}\in{\mathbb C}^{N\times 1}$, ${\bf x}^{\mathrm H}{\bf U}{\bf x}=\sum_{k=1}^K|\varepsilon_k|^2|{\bf h}_k^{\mathrm H}{\bf x}|^2\ge 0$ always holds, $f_{2b}(\bm{\theta})$ is a quadratic concave function of $\bm\theta$. However, when considering the quantization in the RIS model, the constraints in (P3) are not convex, which is difficult to cope with. In our experiments, we may first compute an unquantized optimization result (i.e., adjust the constraints to $|\theta_n| = 1$) and then project it into the corresponding nearest bit. Moreover, the authors in \cite{guo2019weighted} point out that even if we relax the constraints to $|\theta_n|\le 1$, the optimization result will be close to $|\theta_n|=1$ in the end, indicating that the two constraint schemes ($|\theta_n| \le 1$ and $|\theta _n| = 1$) are almost equivalent in this condition. In this perspective, to solve (P3), we develop the following steps:
\begin{itemize}
\item First, solve a convex quadratically constrained quadratic program (QCQP) problem (P4) as follows
\begin{equation}
\text{(P4)}\quad
\begin{aligned}
&\max_{\bm{\theta}}& f_{2b}&(\bm{\theta})\\
&\mathrm{s.t.}& \theta_n&\in \mathcal{F}_c \quad \forall n=1,\cdots,N,
\end{aligned}
\end{equation}
where ${\mathcal {F}}_c=\left\{\theta_n\big | |\theta_n|\le 1,n=1,\cdots,N\right \}$ denotes the set of possible values of phase shifting under flexible control. Define ${\bf e}_i=[0, \cdots, 0, \mathop{1}\limits_{i_{\mathrm{th}}},0, \cdots, 0]$, and $\bf{T}_i = \text{diag}({\bf e}_i)$ is positive definite, thus $\mathcal{F}_c$ is equivalent to $N$ independent constraints:
\begin{equation}
\bm{\theta}^{\mathrm{H}}{\bf T}_i{\bm\theta}\le 1, \quad\forall i=1,\cdots,N.
\end{equation}
It is clear that the feasible set $\mathcal{F}_c$ is convex, thus the problem (P4) is convex. The solvers in convex optimization (e.g., CVX toolbox) can obtain the optimal solutions $\bm{\theta}^{\mathcal{F}_c}$ which obeys the rule that the modulus is close to one. There may be problems of high computational complexity without considering any preconditions of the original problems in the solving process of solvers. To address this issue, we propose a workflow below utilizing the low-rank properties to reduce complexity.
\item Second, project all reflection coefficients to discrete bits according to the rules of closest point projection, i.e.,
\begin{equation}
\theta_n^{\mathcal{F}_d}=\arg \min_{\phi_n\in{\mathcal F}_d} \left |\phi_n-{\theta_n^{{\mathcal F}_c}}\right |.
\label{CPP}
\end{equation}
\end{itemize}
The matrix ${\bf U}\in{\mathbb C}^{N\times N}$ is the sum of $K$ rank-1 matrices, and thus satisfies $\text{rank}({\bf U})\le K\ll N$. Perform Eigenvalue Decomposition (EVD) on $\bf U$, i.e.,
\begin{equation}
{\bf U} = {\bf P}^{\mathrm H}{\bf D}{\bf P},
\end{equation}
where $\bf D$ is a diagonal matrix. Denote the entries on its diagonal by ${\bf d} = \begin{bmatrix}d_1&d_2&\cdots&d_K&0&\cdots\end{bmatrix}$. The objective function is transformed into 
\begin{equation}
f_{2b}({\bm \theta}) = -{\bm\theta}^{\mathrm T}{\bf P}^{\mathrm H}{\bf DP}{\bm \theta}^{\ast}+2\Re{({\bm\theta}^{\mathrm{T}}{\bf v})}+C.
\end{equation}
Denote the new optimization variables as 
\begin{equation}
{\bm \omega}={\bf P}{\bm\theta}^{\ast}.
\label{new varia}
\end{equation}
Since the unitary matrix does not affect the constraints, the optimization problem is further transformed into 
\begin{equation}
\begin{aligned}
&\max_{\bm\omega} &f_{2c}({\bm\omega}) &= -{\bm\omega}^{\mathrm H}{\bf D}{\bm \omega}+2\Re{({\bm\omega}^{\mathrm H}{\bf Pv})}+C \\
&\text{s.t.}&\omega_n&\in{\mathcal{F}}_c.
\end{aligned}
\label{omega}
\end{equation}
Since the matrix $\bf D$ is diagonal, the term ${\bm\omega^{\mathrm H}}{\bf D}{\bm\omega}$ in (\ref{omega}) can be rewritten as 
\begin{equation}
{\bm\omega}^{\mathrm H}{\bf D}{\bm\omega} = \sum_{i=1}^N \omega_i^{\ast} d_i\omega_i=\sum_{i=1}^K\omega_i^{\ast} d_i\omega_i.
\end{equation}
Denote ${\bf Pv}$ by $\bf b$ and the term $2\Re{({\bm\omega}^{\mathrm H}{\bf Pv})}$ is represented by
\begin{equation}
2\Re{({\bm\omega}^{\mathrm H}{\bf Pv})}=2\Re{(\sum_{i=1}^N\omega_i^\ast b_i)}.
\end{equation}
By letting ${\partial f_{2c}(\bm\omega)}/{\partial\omega_k}=0$, the solution to (\ref{omega}) can be obtained. It is important to highlight that matrix $\bf P$ is formed by eigenvectors, and the vector $\bf v$ is approximately orthogonal to most of the column vectors of $\bf P$. This signifies that only the first $K$ terms in vector $\bf b$ have non-zero values. Thus the closed-form solution to (\ref{omega}) is
\begin{equation}
    \omega_k=\left \{\begin{aligned}
        &\frac{2\Re({b_k})}{d_k}, &\quad k = 1,\ldots,K, \\
        & 0, &\quad k = K+1, \ldots, N.
    \end{aligned}\right .
\end{equation}
With this characteristic, we exclusively calculate the $K$ eigenvectors and their corresponding eigenvalues, avoiding the need for a complete EVD and potentially decreasing overall complexity. The computed reflection coefficients are obtained according to (\ref{new varia}). The final step is the closest point projection, which concludes the flow of a single iteration. Denote the result in the $i$-th iteration by $\bm{\theta}^i$, and the updating flag is whether the objective function of (P3) is increasing, which also guarantees the non-decreasing property of the objective function of (P1). The procedures of this algorithm are summarized in Algorithm~\ref{alg:1}. The total computational complexity of EM-GAMP algorithm to recover $K$ vectors with dimension $N$ is $\mathcal{O}(KN\log (N))$. In each iteration, the computational complexity is determined by four steps. The complexities of step 3 (updating the value of $\bm\alpha$) and step 4 (updating the value of $\bm\varepsilon$) are both $\mathcal{O}(KN)$, while the complexity of step 5 is dominated by EVD, which is $\mathcal{O}(N^3)$ at most. The projection in step 6 has a complexity of $\mathcal{O}(N)$. Additionally, in our simulations, the algorithm consistently converges within approximately $10$ iterations, justifying the choice of $t_{\text{max}} = 10$. Given that $K \ll N$ commonly holds in RIS model, the overall time complexity of the QTLM algorithm is $\mathcal{O}(t_{\text{max}}N^3)$.
We will apply this algorithm in our experiments to obtain the near optimal reflection coefficients to configure the RIS. 

\begin{algorithm}[ht]
  \renewcommand{\algorithmicrequire}{\textbf{Input:}}
  \renewcommand{\algorithmicensure}{\textbf{Output:}}
  \caption{Proposed Quadratic Transform-based Low-rank Multi-user Beamforming Algorithm (QTLM)}
  \label{alg:1}
  \begin{algorithmic}[1]
    \REQUIRE Randomly generated initial reflection coefficients vector $\bm\theta^0\in{\mathcal{F}_d}$, the maximum number of iterations $t_{\text{max}}$, the direct link channel coefficient by LS method and the cascaded channel vector estimated by solving (\ref{CS_constr}).
    \ENSURE Computed reflection coefficients $\bm{\theta}^{\ast}$.
    \STATE $\bm{\theta}^{\ast}\leftarrow \bm{\theta}^0$
    \FORALL{$t = 1,\cdots,t_{\text{max}}$}
    \STATE Update $\bm{\alpha}^i$ according to (\ref{alpha}) and (\ref{gamma});
    \STATE Update $\bm{\varepsilon}^i$ according to (\ref{varepsilon_opt}) and calculated $\bm{\alpha}^i$;
    \STATE Calculate a continuous result by solving (P4) following the procedures of closed-form equations between (34) and (39);
    \STATE Project the continuous results into quantized one according to (\ref{CPP}), denoted by $\bm{\theta}^i$;
    \IF {$f_{2b}(\bm{\theta}^i)>f_{2b}(\bm{\theta}^\ast)$}
      \STATE $\bm{\theta}^{\ast}\leftarrow \bm{\theta}^i$;
    \ELSE
      \STATE \textbf{break};
    \ENDIF
    \ENDFOR
    \STATE \textbf{return} The reflection coefficients $\bm{\theta}^{\ast}$.
  \end{algorithmic}
  \label{BF Alg}
\end{algorithm}

\section{Numerical Results}
\label{sec 5}
In this section, we present numerical results to demonstrate the effectiveness of the proposed QTLM scheme in the RIS-aided multi-user communication systems.

\subsection{Simulation Settings}
\begin{figure}
\includegraphics[width = \linewidth]{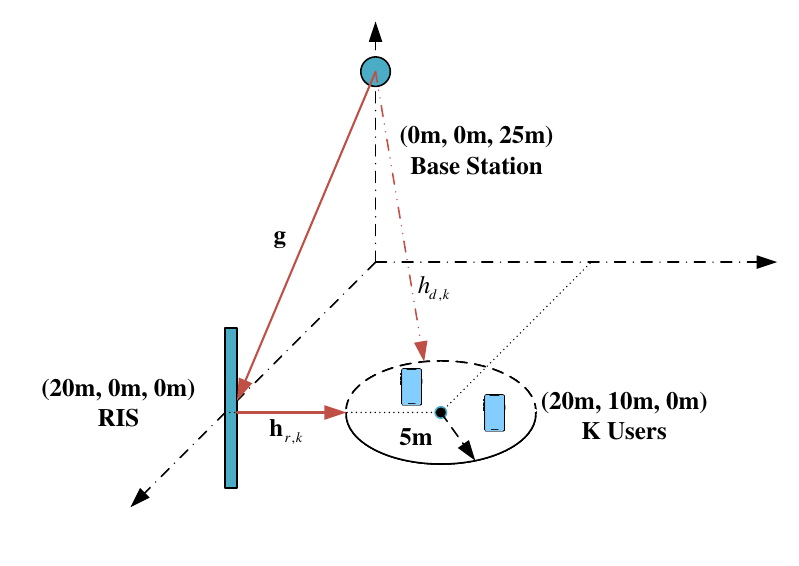}
\caption{The simulated RIS-aided $K$-users communication system comprising of one base station and one $N$-element RIS.}
\label{simuPic}
\end{figure}

We adopt a three-dimensional coordinate configuration, where the locations of the BS and the passive RIS are set to $(0,0,25)$ meters and $(20,0,0)$ meters, respectively, and $K$ users are uniformly distributed in a circle located in the horizontal plane, with $(20,10,0)$ meters as the center and a radius of $5$ meters. The geometric location of these components is shown schematically in Fig. \ref{simuPic}. All the angle information in channel model is determined according to the relative locations. Each link in ${\bf h}_{r,k}$ and ${\bf g}$ is subject to both path loss and small-scale fading. The path loss model is set according to 3GPP propagation environment described in \cite{3GPP2019Study}. For small-scale fading, we employ the standard Rician channel model, assigning Rician factor of $10\text{ dB}$ to ${\bf h}_{r,k}$ and ${\bf g}$. The Power Spectral Density (PSD) of the white noise is set to $-170\text{ dBm/Hz}$, and the bandwidth of the simulated communication system is set to $2\text{ MHz}$. The RIS element is quantized to 1-bit. For clarity, the system parameters are summarized in TABLE \ref{simulation table}. 
\begin{table}[htb!]
\renewcommand{\arraystretch}{1.2}
\centering
\caption{Simulation Setups}
\label{simulation table}
\begin{tabular}{|c|c|}
\hline
\textbf{Parameter}      		& 	 \textbf{Value} 									\\ \hline
Base station location 			&    $(0,0,25)\text{ m}$       						    \\ \hline
RIS central location  			&    $(20,0,0)\text{ m}$       			 			    \\ \hline
Users central location			&    $(20,10,0)\text{ m}$								\\ \hline
Dimension of RIS 				&    $16\times 32$										\\ \hline
Number of users					&    $K=2,4$												\\ \hline
PSD of noise          			&    $-170\text{ dBm/Hz}$         						\\ \hline
Path loss         		        &    $28 + 22\log_{10}(d) + 20\log_{10}(f_c)$   		\\ \hline
Rician factor                   &    $10\text{ dB}$    									\\ \hline
Center frequency 				&    $f_c = 5.8\text{ GHz}$ 							\\ \hline
\end{tabular}
\end{table}

In light of the necessity for real-world deployment, we excluded algorithms with high complexity from our simulations. Therefore, for the purpose of comparison, we simulated the performance of the following existing schemes:
\begin{itemize}
\item Genetic algorithm based scheme (GA) \cite{zhi2022power}. Genetic algorithms serve as powerful tools in the realm of optimization and search problems, drawing inspiration from the intricate mechanisms of biological evolution. Given this characteristic, genetic algorithms prove effective for addressing beamforming problems of RIS falling under the category of discrete optimization, showcasing outstanding performance in simulations \cite{peng2021analysis}. During the implementation of this algorithm, the objective was set as the spectral efficiency of the entire multi-user system.
\item Greedy fast beamforming algorithm (GFBA) \cite{pei2021ris}. It mainly utilizes the dominant paths between BS/UE and the RIS, and formulates the reflection coefficients as  a column vector in the DFT matrix of a specific dimension, which could greatly reduce the training/feedback overhead when configuring the surface. This algorithm has been proven to be highly effective in practical single-user communication environments.
\item Random beamforming (RBF). The fundamental concept of this scheme is to choose a reflection coefficient vector which obtains the best gain of the  performance indicator from a pre-generated random set. The critical parameter for this algorithm is the size of the set.
\end{itemize}
\subsection{Simulation Results}
The detailed key parameters for each algorithm are summarized in TABLE \ref{paraSetup}. To illustrate the performance robustness of our algorithm for generating channels with different user locations, all the simulation results are averaged over $10^4$ channel realizations. More precisely, we randomly generate the locations of $K$ users for $100$ snapshots and conduct Monte Carlo simulations $100$ times for each snapshot to obtain the results. 

\begin{table}[htb!]
\caption{Parameters setup for different algorithms}
\centering
\label{paraSetup}
\renewcommand{\arraystretch}{1.3}
\resizebox{\linewidth}{!}{
\begin{tabular}{|c|c|c|}
\hline
\textbf{Algorithm}         & \textbf{Parameter}                                 & \textbf{Value}         							 \\ \hline
\multicolumn{1}{|c|}{QTLM} & \multicolumn{1}{c|}{Number of samples} & \multicolumn{1}{c|}{$P=200$}   	 							 \\ \hline
\multirow{4}{*}{GA}        & Population size                                    & $1000$             							 \\ \cline{2-3} 
                           & Maximum number of generations                      & $80$                  							 \\ \cline{2-3} 
                           & Mutation rate                                      & $0.2$                  							 \\ \cline{2-3} 
                           & Crossover rate                                     & $0.5$                  							 \\ \hline
\multirow{2}{*}{GFBA}      & Maximum number of iterations                       & $10$                   							 \\ \cline{2-3} 
                           & Initialization configuration                       & \makecell[c]{Homogeneous \\ state}                 \\ \hline
\multicolumn{1}{|c|}{RBF}  & \multicolumn{1}{c|}{Size of set} & \multicolumn{1}{c|}{51200}   		 	 						 	 \\ \hline
\end{tabular}}
\end{table}

\begin{figure}
    \centering
    \includegraphics[width = \linewidth]{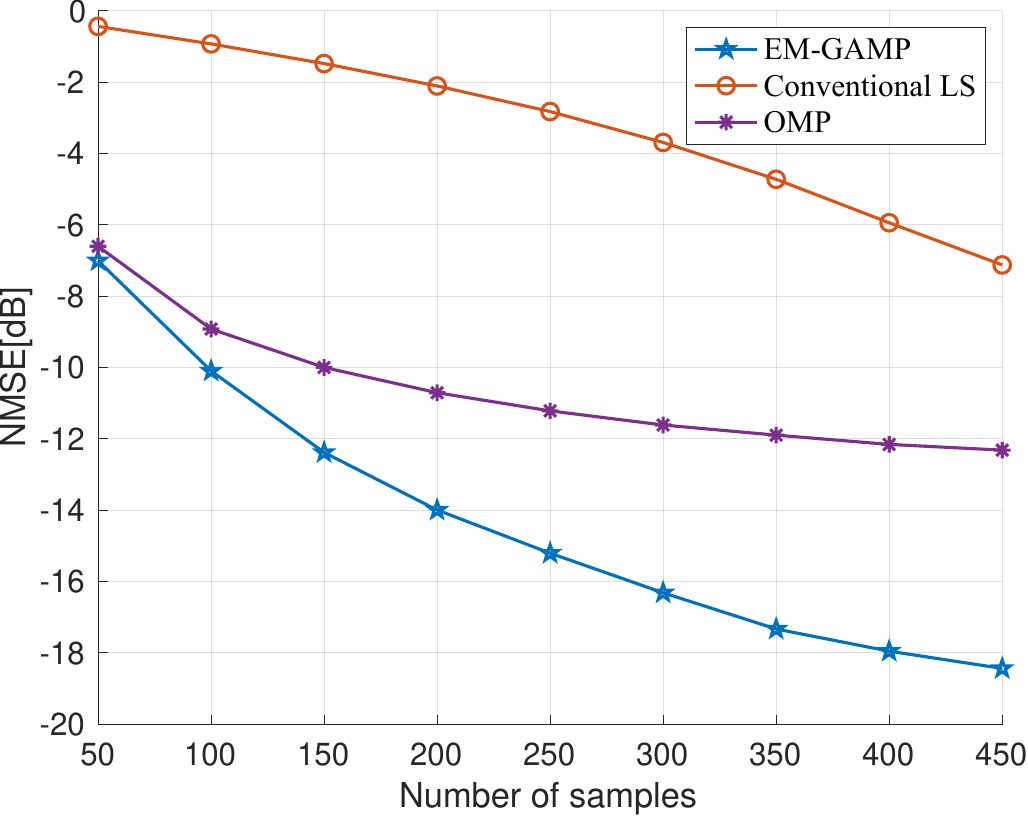}
    \caption{NMSE of the cascaded channel vs. the number of samples utilized in the channel estimation phase.}
    \label{CEFig}
\end{figure}

We first demonstrate the performances of the channel estimation algorithms. In Fig. \ref{CEFig}, a comparison of Normalized Mean Square Error (NMSE) against the number of samples used in the channel estimation phase is presented. It is evident that the EM-GAMP algorithm outperforms other conventional reconstruction methods. Furthermore, when an ample number of samples are available, the NMSE performance tends to be converged. This trend is consistently observed in subsequent experiments.

\begin{figure}
\centering
\subfigure[$K=2$.]{\includegraphics[width = \linewidth]{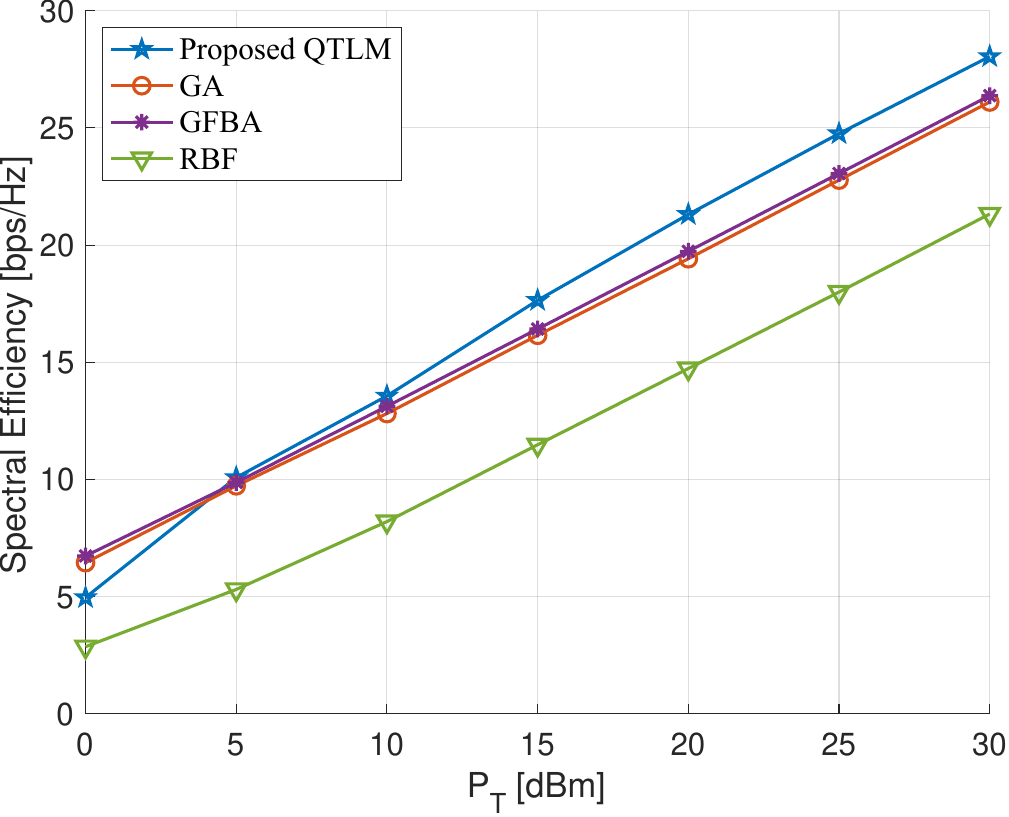}}
\subfigure[$K=4$.]{\includegraphics[width = \linewidth]{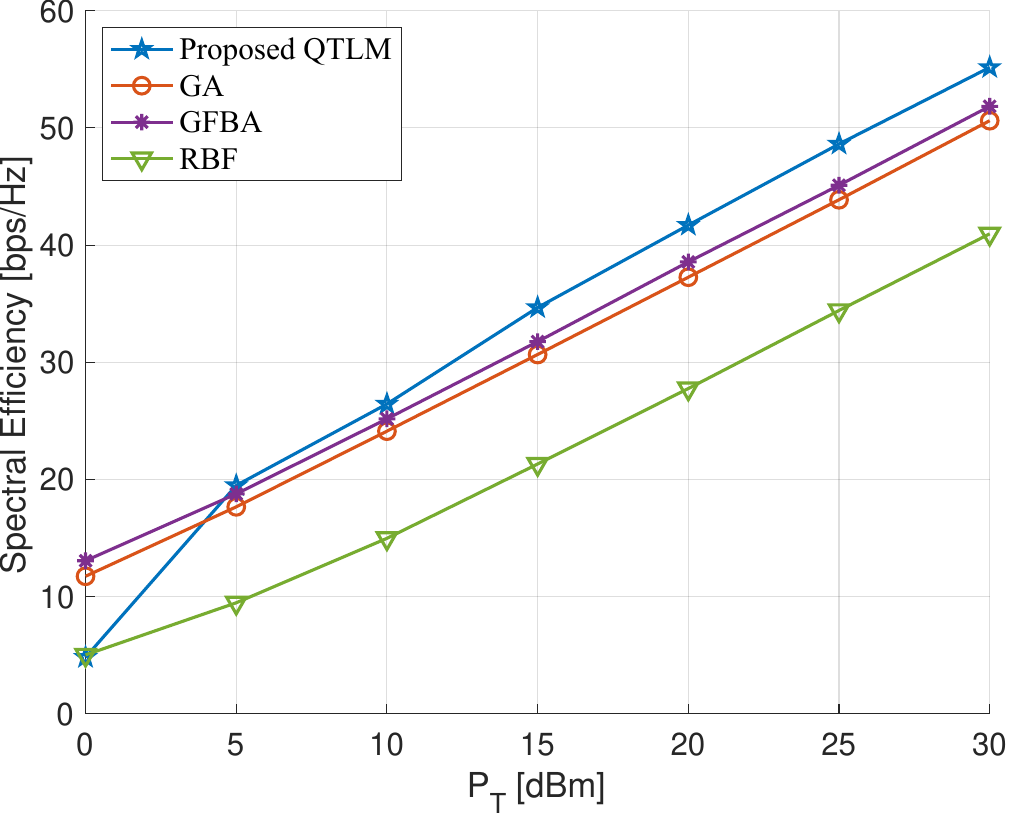}}
\caption{Achievable spectral efficiencies vs. the transmitting power in BS. (a) $K=2$. (b) $K=4$.}
\label{simRes}
\end{figure}

The spectral efficiencies achieved by different schemes under various transmitting powers are illustrated in Fig. \ref{simRes}. As the transmitting power increases, the performance advantage of the proposed QTLM algorithm compared to other algorithms becomes more pronounced, fundamentally stemming from the improved accuracy in channel estimation. Besides, the QTLM algorithm is able to reduce the pilot length by about $60\%$ while maintaining excellent performance. This also indirectly illustrates the practical significance of the QTLM algorithm.
\section{Experimental Results}
\label{sec 6}

\subsection{Experimental Scenarios and Test Process}
Our RIS-aided wireless communication system is composed of host Personal Computers (PCs), Universal Software Radio Peripherals (USRPs), RIS, and a control board. The whole system works at $5.8 \text{ GHz}$. The RIS board is composed of $512$ elements, with $16$ rows and $32$ columns, each of which is quantized to 1-bit and designed the same as \cite{pei2021ris}. USRPs are used to process signal, and the control board is carefully designed to adjust the biased voltage of every individual element of RIS. The detailed parameters of hardware are summarized in TABLE \ref{hardware_config}.
\begin{table}[htb!]
\caption{\uppercase{Detailed Information on Hardware Modules}}
\label{hardware_config}
\renewcommand{\arraystretch}{1.3}
\resizebox{\linewidth}{!}{
\begin{tabular}{|cc|c|}
\hline
\multicolumn{2}{|c|}{\textbf{Hardware configuration}}                          & \textbf{Value}                       \\ \hline
\multicolumn{1}{|c|}{\multirow{5}{*}{RIS element}}  & Polarization             & Unipolar                             \\ \cline{2-3} 
\multicolumn{1}{|c|}{}                              & Frequency                & $5.8$ GHz                            \\ \cline{2-3} 
\multicolumn{1}{|c|}{}                              & Quantization             & $\tau=1$                             \\ \cline{2-3} 
\multicolumn{1}{|c|}{}                              & \multirow{2}{*}{Spacing} & $d_y = 14.3\text{ mm}$               \\ 
\multicolumn{1}{|c|}{}                              &                          & $d_z = 10.27\text{ mm}$              \\ \hline
\multicolumn{1}{|c|}{\multirow{2}{*}{RIS board}}    & Number of elements       & $16 \times 32$                       \\ \cline{2-3}
\multicolumn{1}{|c|}{}                              & Control signal           & Individual                           \\ \hline
\multicolumn{1}{|c|}{\multirow{3}{*}{Horn antenna}} & Gain                     & $17.1 \text{ dBi @ }5.8\text{ GHz} $ \\ \cline{2-3} 
\multicolumn{1}{|c|}{}                              & Aperture                 & $169$ mm$\times$ $119$ mm            \\ \cline{2-3} 
\multicolumn{1}{|c|}{}                              & Beamwidth                & $30^\circ$                           \\ \hline
\end{tabular}}
\end{table}

\begin{figure}[htbp]
\centering
\subfigure[]{\includegraphics[width = \linewidth]{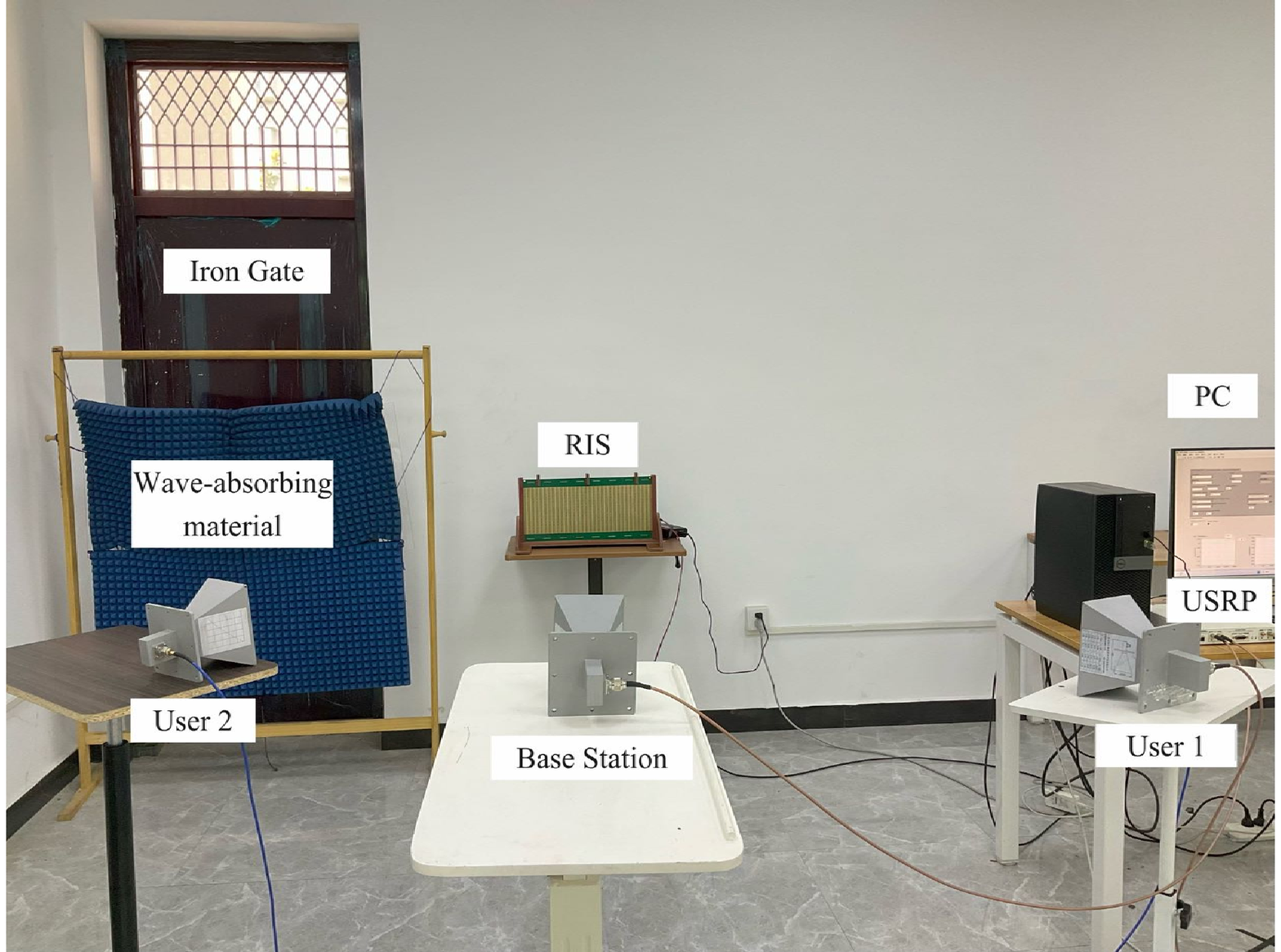}}
\subfigure[]{\includegraphics[width = \linewidth]{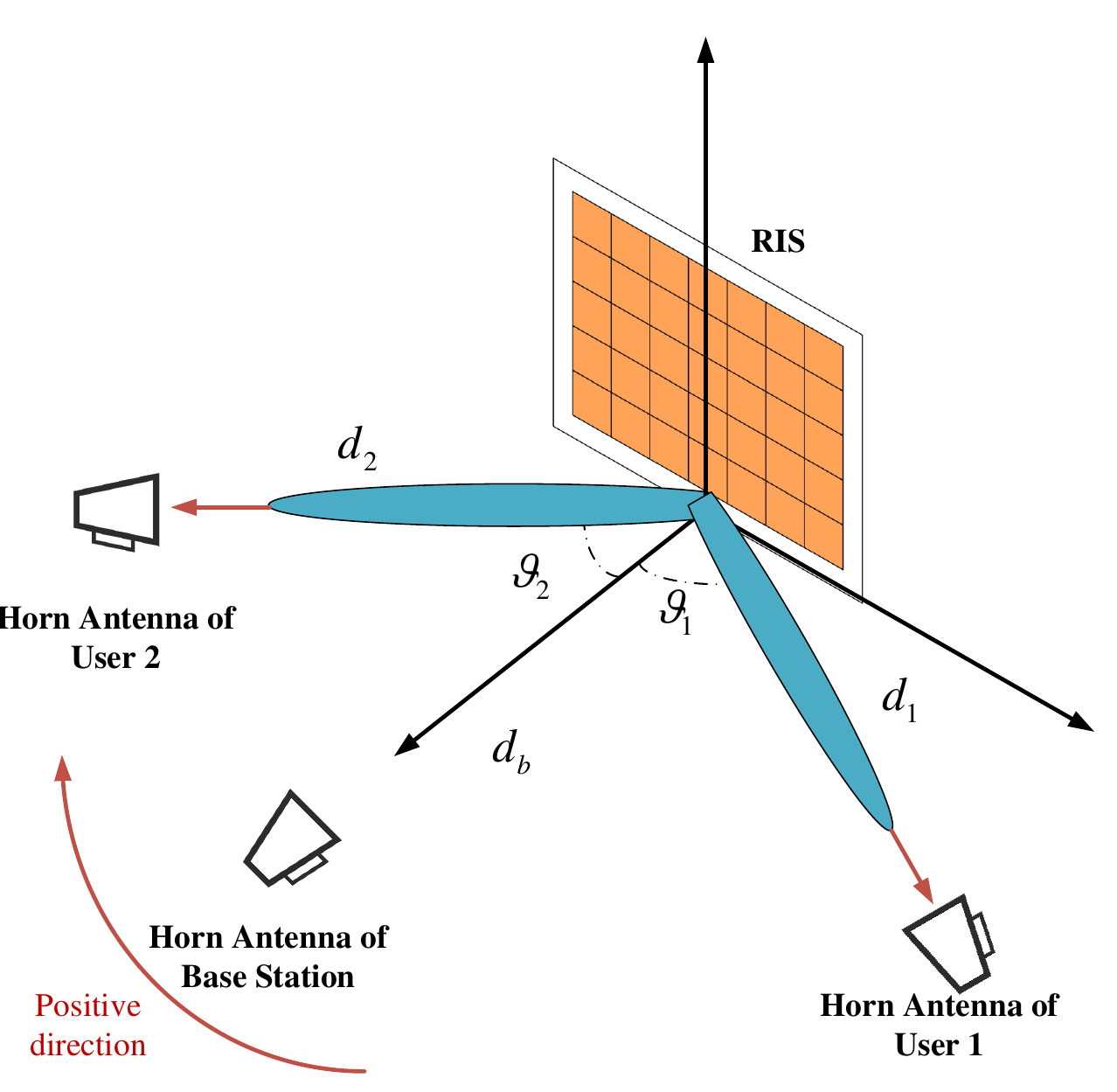}}
\caption{Indoor test scenario. (a) Real-world environment, (b) Location schematic.} 
\label{labexp}
\end{figure}

Our experiments are made in indoor environment shown in Fig. \ref{labexp}. The RIS board is positioned against the wall, and the antennas of the base station and users are placed on one side of RIS and on the same horizontal plane. The transmitting horn antenna of the base station is perpendicular to the RIS, i.e., the angle of incidence is zeros. The distance between the  antennas of the base station and the RIS, the $k$-th user and the RIS are denoted by $d_b$ and $d_k$ respectively. The antennas of users receive electromagnetic signals obliquely emitted by the RIS. Denote the angle between the receiving antenna of the $k$-th user and the center of the RIS by $\theta_k$ (let clockwise rotation be positive, and counterclockwise negative). The wave-absorbing material is placed at the left rear mainly to prevent the reflection from the iron gate.

According to the algorithm flow summarized in Sec. \ref{sec 3} and Sec. \ref{sec 4}, our testing process will be divided into the following steps:

\begin{enumerate}[(1)]
\item  Initialization: Randomly generate reflection coefficients matrix following Rademacher distribution.
\item  Setting: Transmit known pilots to all users.
\item  Direct link estimation: Traditional LS channel estimation of direct link after turning off the RIS.
\item  RIS-aided link estimation and reflection coefficients calculation: Turn on the RIS and switch the reflection coefficients of the RIS depending on the pre-generated Rademacher matrix, and at the same time the users sample received signal and reconstruct the cascaded channel vector based on the samples. With the objective of maximizing the spectral efficiency of the whole system, utilize the QTLM algorithm and calculate the reflection coefficients.
\item  Beamforming: Apply the computed reflection coefficients to the RIS.
\end{enumerate}

It is worth noting that, the estimation of the direct link can be solved by conventional LS method once we assume that the RIS absorbs the incident electromagnetic wave, which was implemented in a dedicated time slot, separate from the estimation of the cascaded channel. In our experiments, in order to simply reach the state of absorption, we cover the RIS with wave-absorbing materials. However, another approach to reach the same effect is illustrated in \cite{imani2020perfect}. The work points out that, by tuning the resistance and reactance of the element, the input impedance of RIS can be changed, which will result in perfect absorption of the incident electromagnetic wave when it is matched with free-space impedance. 

\subsection{Channel Estimation Algorithm Experiments}
For a single user RIS-aided communication system, the optimization problem is reduced from (P1) to
\begin{equation}
\begin{aligned}
&\max_{\bm{\theta}} \quad &\gamma&=\frac{\left |\hbar_d+{\bm{\theta}^{\mathrm T}{\bf h}}\right |^2}{\sigma^2}\\
&\mathrm{s.t.}\quad &\theta_n&\in{\mathcal{F}_d}\quad \forall n=1,\cdots,N.
\end{aligned}
\end{equation}
Once we obtain the channel state information of RIS-aided link, i.e., the cascaded channel ${\bf h}$, the optimal reflection coefficients of RIS is approximately expressed as \cite{wu2019intelligent}
\begin{equation}
\bm{\theta}^{\text{SU}} = \frac{\hbar_d}{|\hbar_d|}\cdot\frac{{\bf h}^{\ast}}{\Vert {\bf h}\Vert},
\end{equation}
which straightforwardly concludes from the combination of Maximum Ratio Transmission (MRT) and phase alignment. If a quantized phase shifting scheme is taken into account, $\bm\theta$ is projected into the closest point from $\bm\theta^{\text{SU}}$ following the guidelines in (\ref{CPP}). Therefore, the performance of the single-user system can reflect the accuracy of channel estimation algorithm, which can be observed by comparing the received power before and after the beamforming of RIS. The performance of the algorithm is reflected by a parameter $G\text{ (dB)}$ defined as 
\begin{equation}
G \text{ (dB)} =10\log_{10} \left (\frac{P_{r}^{\text{B}}}{P_{r}^{\text{O}}}\right ),
\label{gain}
\end{equation}
where $P_r^{\text{B}}$ is the received power after configuring the beamforming of RIS and $P_r^{\text{O}}$ is the received power when the RIS is powered off, which are both calculated from LabVIEW and follow 
\begin{equation}
P_{r}\text{ (W)} = \frac{1}{T}\sum_{t=1}^T |x_t|^2,
\end{equation}
where $x_t$ is the sample in the received frames, and $T$ is the number of samples contained in a baseband symbol. Besides, when the relative locations of the user, the base station, and the RIS are fixed, the performance of the CS reconstructing algorithm suffers from influences of number of samples and signal power. Thus, we conduct two relevant types of experiments to verify the effectiveness of the proposed channel estimation algorithm.

\begin{figure}[htbp]
\centering
\includegraphics[width = \linewidth]{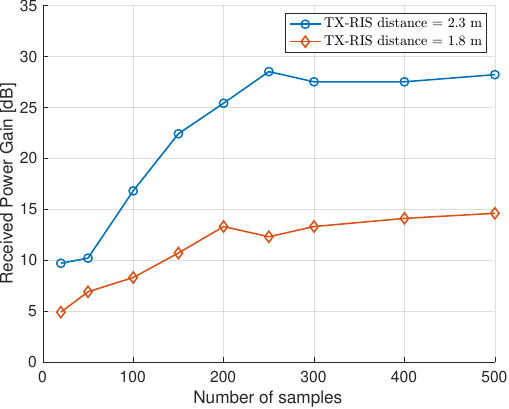}
\caption{Received power gain vs. number of samples in the single-user scenario, experiments settings are $d_1 = 2.50\text{ m}, \vartheta_1 = 25^\circ$.}
\label{samNumCE}
\end{figure}

The boundary of the near field and far field in our RIS model is around $B\simeq 2.9\text{ m}$ according to (\ref{bound}). However, since the sparsity of angular channel model is not satisfied in the near field, the received power gain is lower compared with the far field scenarios. This effect is illustrated in Fig. \ref{samNumCE}. The average peak received power gain of the algorithm is $13.5\text{ dB}$ when the distance is $1.8\text{ m}$, while the average peak received power gain is $27.9\text{ dB}$ when the distance is $2.3\text{ m}$.

\begin{figure}[htbp]
\centering
\includegraphics[width = \linewidth]{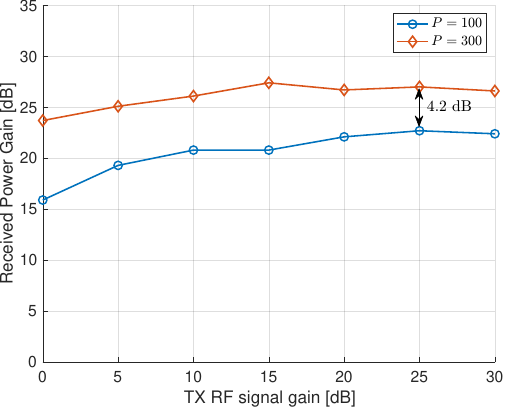}
\caption{Received power gain vs. TX RF signal gain in the single-user scenario, experiments settings are $\text{TX-RIS distance} = 2.30\text{ m}, d_1 = 2.60\text{ m}, \vartheta_1 = 22^\circ$.}
\label{GainTXCE}
\end{figure}
By changing the dimension of the sensing matrix in (\ref{compressed_sensing}) when randomly generated, the algorithm utilizes different samples to recover the channel vector. It is also observed in Fig. \ref{samNumCE} that, as the number of samples increases, the received power gain of the single user grows gradually and the peak received power gain compared with the ``OFF" state of RIS is $28.2\text{ dB}$ in such a single-user scenario. The curve tends to converge between $200$ and $250$ samples. Therefore, augmenting the number of samples beyond this range does not bring a performance enhancement for the channel estimation algorithm. 

In order to observe the effect of signal power on algorithm performance, we conducted experiments in two strategies for the number of samples, i.e., $P=100$ and $P=300$, which corresponds to two curves in Fig. \ref{GainTXCE}. To ensure the accuracy of the experiment, the measurement matrix is identical when the number of samples is the same. The increase of the signal power generally results in a performance gain, which is in line with the intuition of the EM-GAMP algorithm. In contrast, increasing the signal power when the number of samples is small is of greater importance. As can be seen from the two curves in the figure, in both cases, a $30\text{ dB}$ increase in transmitting signal power brings a $2.9 \text{ dB}$ and $6.5 \text{ dB}$ gain for the performance of the algorithm, respectively. There are also differences in the upper limits of the estimated performance of the measurement matrix for different dimensions, which is reflected by the received power gain gap of approximately $4.2\text{ dB}$ in Fig. \ref{GainTXCE}.

\subsection{Multi-user Passive Beamforming Experiments}
\subsubsection{Correction on noise power}
In our passive beamforming algorithm, the noise power is needed at each iteration, i.e., the updating value of $\bm\alpha$ in (\ref{alpha}). Its accuracy will directly affect the algorithm performance. We find that the noise floor in RF devices is insufficient for a complete measure of noise power, making the algorithm fail to compute the proper reflection coefficients results. Therefore, we take the defects in the RF chains into account, which is a different form of noise from Gaussian white noise and includes the influences of phase noise, noise figure, I/Q phase imbalance, and power supply noise. All these effects in the receiver can be reflected in the parameter of Receive Modulation Error Ratio (RxMER). One can calculate the average distance between the reference signal vector and measured signal vector in the constellation as RxMER in the following equation
\begin{equation}
\text{RxMER} = \frac{\sum_{i=1}^N|R_i|^2}{\sum_{i=1}^N\left |S_i-R_i\right |^2},
\end{equation}
where $N$ is the number of samples in frames, $S_i$ is the measured signal sample, and $R_i$ is the reference signal sample. In this background, RxMER is a key indicator to quantify the transmitting performance of a wireless communication system, which measures Gaussian noise and other uncorrectable impairments of the received constellation\cite{hranac2006digital}. 

Note that RxMER and SNR are equivalent when only Gaussian noise is present in the system. In other words, RxMER is a special manifestation of SNR. In the experiments, we can compute the RxMER value to substitute the noise floor part, which is equivalent to a correction for noise power. The correction value of noise power is 
\begin{equation}
\bar{\sigma}^2 \text{ (W)}= \frac{P_{r,k}}{\text{RxMER}},
\label{RxMER noise}
\end{equation}
where $P_{r,k}$ is the average received power of the $k$-th user calculated in LabVIEW. In the later process of experiments, the noise correction value in (\ref{RxMER noise}) will be used instead of the noise floor for the calculation, making it as close to the true value as possible.

\subsubsection{Spectral efficiency of two users}
To illustrate the performance of the multi-user passive beamforming algorithm, we set up two users to conduct experiments. The relative location of all the transceivers is $(d_b,d_1,d_2,\theta_1,\theta_2)=(2.34\text{ m},2.26\text{ m},2.00\text{ m},-28^\circ,21^\circ)$. Since our objective is to maximize the spectral efficiency of the whole communication system for a multi-user situation, the effectiveness of the algorithm is reflected by the spectral efficiency defined as 
\begin{equation}
\text{S (bps/Hz)}=\sum_{k=1}^K\log_2(1+\frac{P_{r,k}}{\sigma^2}),
\end{equation}
where $P_{r,k}$ is the average received power of the $k$-th user after beamforming of the RIS. 

\begin{figure}[htb]
\centering
\includegraphics[width = \linewidth]{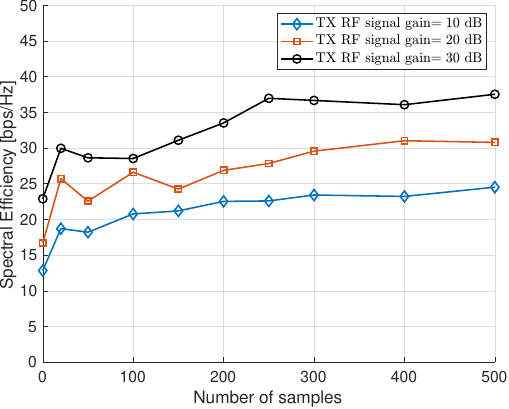}
\caption{The spectral efficiency after configuring reflection coefficients of RIS under the different numbers of samples and TX RF signal gains, the points with the number of samples of $0$ indicate the spectral efficiencies when the RIS is powered off.}
\label{multi-user_sum_rate}
\end{figure}

Similarly, we change experimental conditions, and a figure of the variation of the spectral efficiency with the number of samples and the TX RF signal gain was obtained (the points in the figure with the number of samples of 0 indicate the spectral efficiencies when the RIS is powered off). Fig. \ref{multi-user_sum_rate} shows that, as the number of samples grows, the spectral efficiency of the whole communication system increases. The saturation point of the system capacity is between 250 and 300, and then increasing the number of samples does not deliver a noticeable performance boost. Increasing the number of samples from 20 to 500 gives the system an average spectral efficiency increase of $6.15\text{ bps/Hz}$. If compared with the situation when the RIS is powered off, the average spectral efficiency gain grows to $13.48 \text{ bps/Hz}$.

\begin{figure}[htb!]
\centering
\subfigure[]{\includegraphics[width = \linewidth]{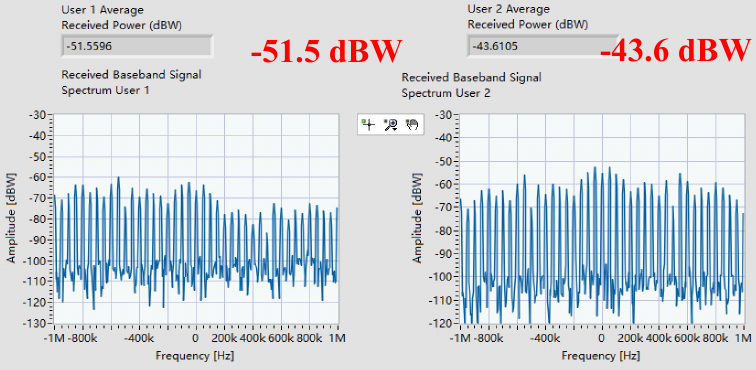}}
\subfigure[]{\includegraphics[width = \linewidth]{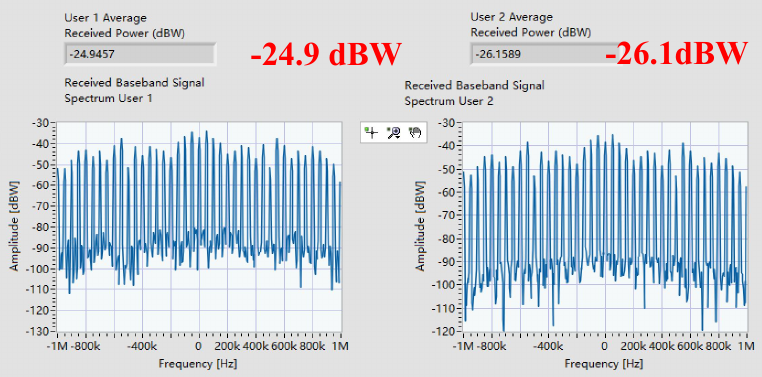}}
\caption{Average received power and spectrum of RX baseband signal for two users obtained from LabVIEW. The experimental results are obtained when $P = 500$, $\text{TX RF signal gain} = 30 \text{ dB}$. (a) All the RIS elements are powered off. (b) After configuring reflection coefficients of the RIS.}
\label{spect}
\end{figure}
The received power gains of each user in different experimental configurations are also recorded in TABLE \ref{gain_multi-user}. Fig. \ref{spect} shows the spectrum and average received power of two users before and after beamforming of the RIS, when the number of samples is $500$ and the TX RF signal gain is $30\text{ dB}$.
The received powers of two users when the RIS is powered off are $-51.5\text{ dBW}$ and $-43.6\text{ dBW}$ respectively. After configuring the reflection coefficients of the RIS, the received power grows to $-24.9\text{ dBW}$ and $-26.1\text{ dBW}$. The two users obtain the gains of $26.6\text{ dB}$ and $17.5 \text{ dB}$ comparing with the powered-off case. The results reveals that the QTLM algorithm makes a decent balance between two users when their respective channel quality is similar, resulting in received power gains of around the same magnitude. However, we also observed that when one of users is in a low-quality environment, the inaccurate channel will interfere with another user in the process of the beamforming algorithm, causing the final throughput rate to not increase as much as expected.

\begin{table}[htb]
\centering
\renewcommand{\arraystretch}{1.2}
\caption{Received Power Gain of each user \\ compared with RIS powered off}
\begin{tabular}{|p{1.28cm}<{\centering}|cc|cc|cc|}
\hline
TX Gain                         & \multicolumn{2}{c|}{$10\text{ dB}$}                                 & \multicolumn{2}{c|}{$20\text{ dB}$}                                 & \multicolumn{2}{c|}{$30\text{ dB}$}                                 \\ \hline
$P$                             & \multicolumn{1}{m{0.7cm}|}{user1} & \multicolumn{1}{m{0.7cm}|}{user2}      & \multicolumn{1}{m{0.7cm}|}{user1} & \multicolumn{1}{m{0.7cm}|}{user2} & \multicolumn{1}{m{0.7cm}|}{user1} & \multicolumn{1}{m{0.7cm}|}{user2} \\ \hline
20                              & \multicolumn{1}{c|}{5.4}   & 12.4                            & \multicolumn{1}{c|}{13.3}  & 13.9                       & \multicolumn{1}{c|}{13.0}  & 8.3                        \\ \hline
50                              & \multicolumn{1}{c|}{6.5}   & 9.8                             & \multicolumn{1}{c|}{2.7}   & 15.0                       & \multicolumn{1}{c|}{5.2}   & 12.1                       \\ \hline
100                             & \multicolumn{1}{c|}{7.7}   & 16.3                            & \multicolumn{1}{c|}{11.3}  & 18.6                       & \multicolumn{1}{c|}{10.5}  & 6.5                        \\ \hline
150                             & \multicolumn{1}{c|}{11.5}  & 13.8                            & \multicolumn{1}{c|}{8.1}   & 14.7                       & \multicolumn{1}{c|}{11.6}  & 13.1                       \\ \hline
200                             & \multicolumn{1}{c|}{12.1}  & 17.2                            & \multicolumn{1}{c|}{16.2}  & 14.5                       & \multicolumn{1}{c|}{19.8}  & 12.2                       \\ \hline
250                             & \multicolumn{1}{c|}{9.1}   & 20.4                            & \multicolumn{1}{c|}{23.0}  & 10.6                       & \multicolumn{1}{c|}{27.0}  & 15.4                       \\ \hline
300                             & \multicolumn{1}{c|}{6.5}   & 17.8                            & \multicolumn{1}{c|}{21.9}  & 16.9                       & \multicolumn{1}{c|}{27.3}  & 14.2                       \\ \hline
400                             & \multicolumn{1}{c|}{13.7}  & 17.7                            & \multicolumn{1}{c|}{23.4}  & 19.8                       & \multicolumn{1}{c|}{25.7}  & 14.0                       \\ \hline
500                             & \multicolumn{1}{c|}{16.7}  & 18.6                            & \multicolumn{1}{c|}{22.8}  & 19.7                       & \multicolumn{1}{c|}{26.6}  & 17.5                       \\ \hline
\end{tabular}
\label{gain_multi-user}
\end{table}

\subsubsection{Radiation Pattern}
\begin{figure}[htbp]
\centering
\includegraphics[width = \linewidth]{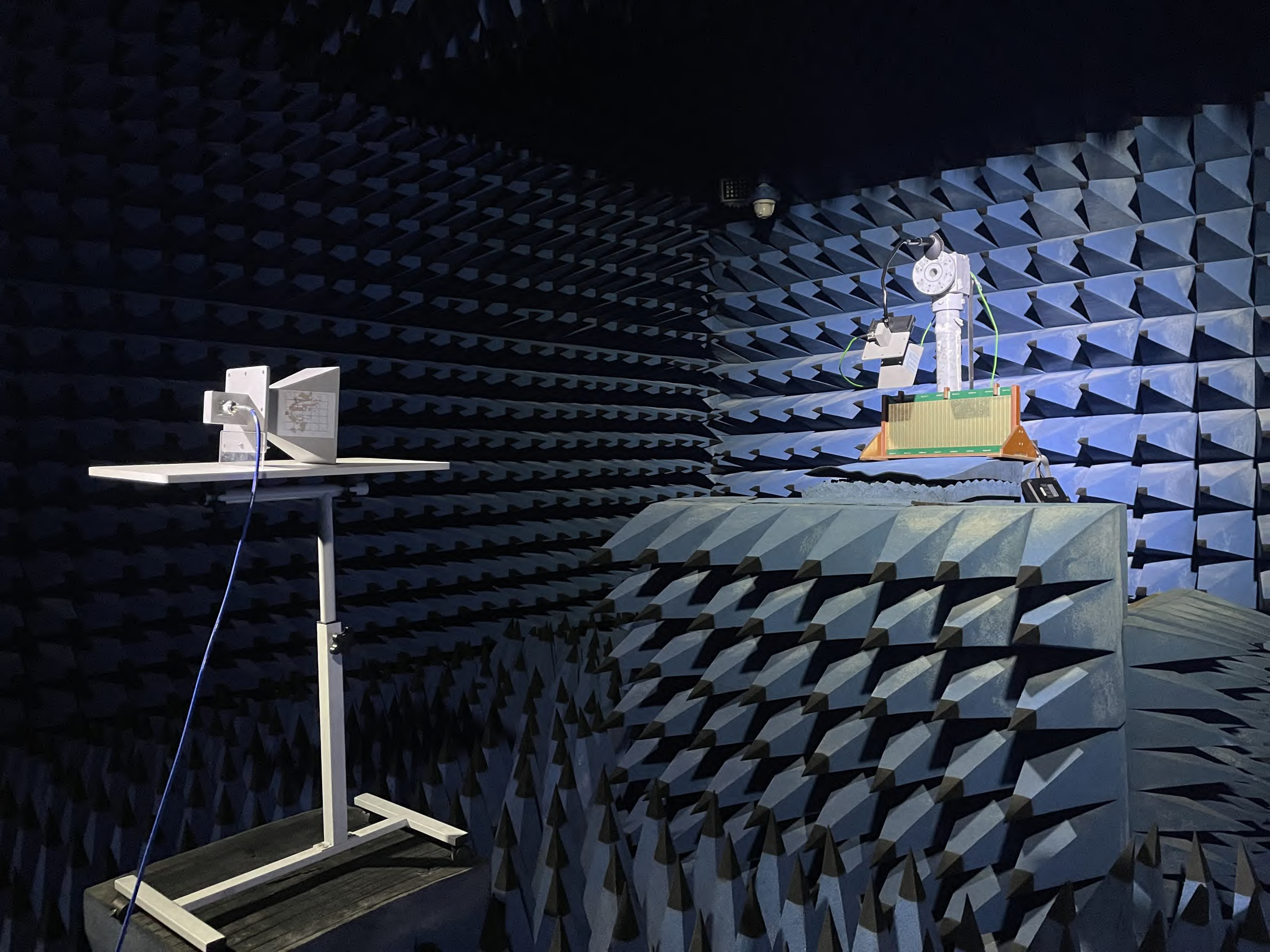}
\caption{The microwave anechoic chamber of size $4\text{ m} \times 6\text{ m} \times 4\text{ m}$, the TX-RIS distance is $0.4\text{ m}$, the RIS-RX distance is $2.2 \text{ m}$. }
\label{chamber}
\end{figure}
For an RIS-aided multi-user communication system, the improvement of spectral efficiency also indicates the multi-beam performance of RIS. Therefore, we conduct radiation pattern experiments in a microwave anechoic chamber as shown in Fig. \ref{chamber}. The transmitting antenna and the RIS are fixed on a rotating platform, while the received antenna is fixed on a table and face the RIS.

\begin{figure}[htbp]
\centering
\includegraphics[width = \linewidth]{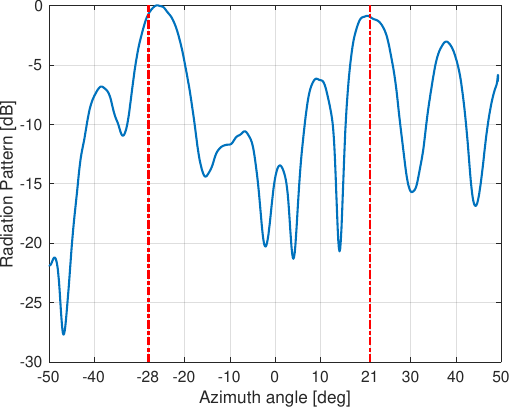}
\caption{Measured normalized radiation pattern of RIS with reflection coefficients obtained from the last section.}   
\label{radPat}
\end{figure}

We used the results obtained from the multi-user experiments conducted above, i.e., the location information is $(d_b,d_1,d_2,\theta_1,\theta_2)=(2.34\text{ m},2.26\text{ m},2.00\text{ m},-28^\circ,21^\circ)$, and measured the radiation pattern of the codeword. The platform rotated and the received powers was recorded in every azimuth direction. The TX-RIS distance is $0.4\text{ m}$, and the RIS-RX distance is $2.2\text{ m}$. The reason that TX-RIS distance does not keep the same as experimental scenario is the length extension limitation of the rotation platform. The radiation pattern of the corresponding codeword is shown in Fig. \ref{radPat}, where the maximum gain is normalized to $0\text{ dB}$. The half-power beamwidth is $9.22^\circ$. It can be observed that there are two wave peaks in radiation pattern, the angle of which are approximately located at $-28^\circ$ and $21^\circ$, which corresponds exactly to our experimental scenario. This pattern implies the multi-beam capability of the RIS board and the multi-user effectiveness of the algorithms.

\section{Conclusion}
\label{sec 7}
In this paper, we proposed a multi-user beamforming scheme for an RIS-aided wireless communication system and implemented all the algorithms in our prototype for the experimental validations. By introducing the angular domain channel model, we solved the channel estimation problem with compressed sensing, and  proposed to generate the measurement matrix with Rademacher distribution to match the 1-bit RIS prototype and utilized the EM-GAMP algorithm to reconstruct the sparse signal. To generate the reflection coefficients of RIS in multi-user scenarios, we proposed a QTLM algorithm, which exploits the structural properties of the channel to enhance convergence speed and reduce complexity. To address RF device imperfections, we put forth a novel method for correcting noise power values, ensuring the efficacy of our algorithms within the RIS prototype. By conducting the experiments with the fabricated RIS prototype system, we verified the effectiveness of all proposed algorithms. The results reveal that the RIS can bring a noticeable increase on the spectral efficiency of the whole multi-user system with our proposed algorithms.  

\appendices

\bibliographystyle{IEEEtran}
\bibliography{IEEEabrv,references.bib}
\begin{IEEEbiography} 
    [{\includegraphics[width=1in,height=1.25in,clip,keepaspectratio]{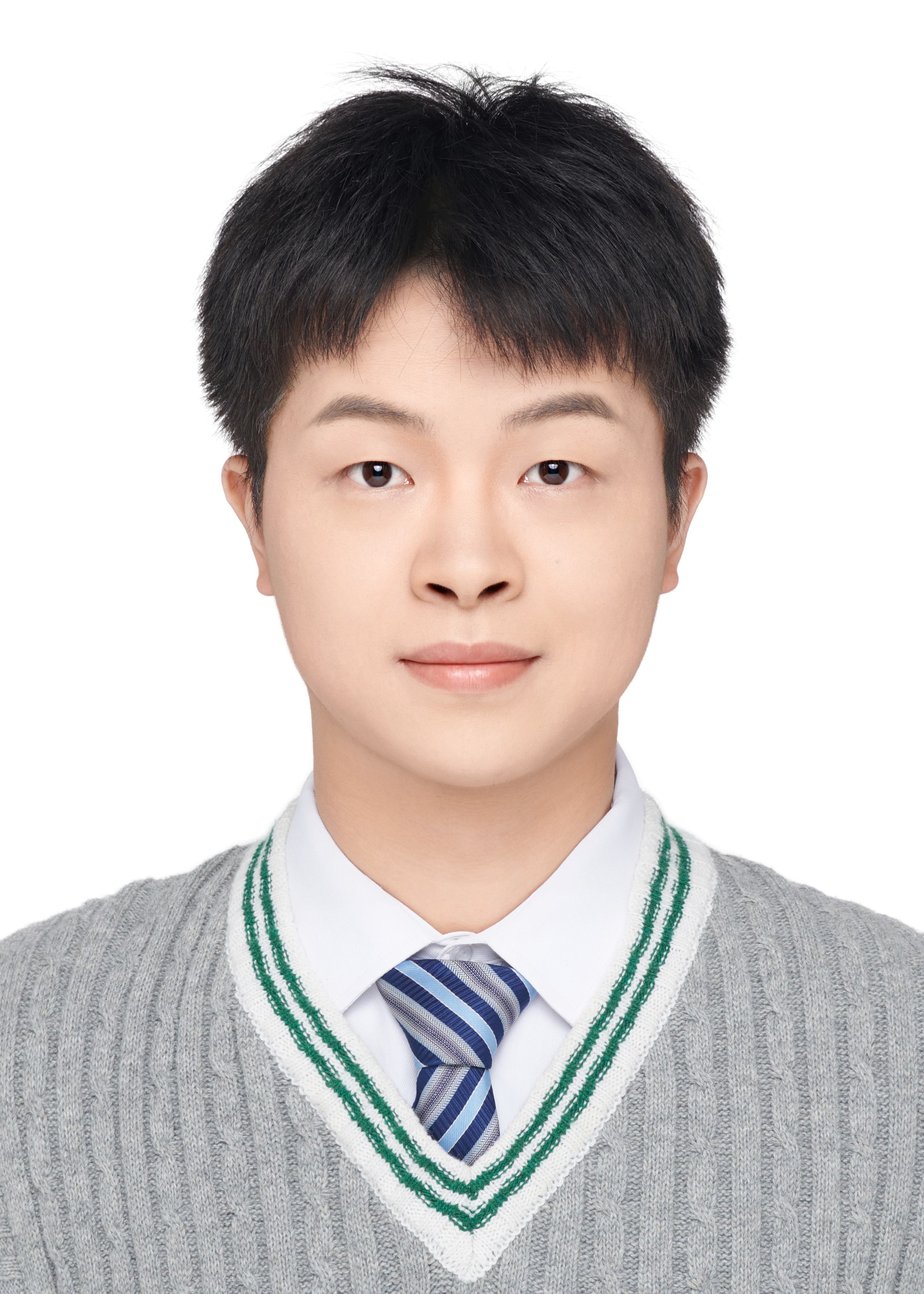}}]{Zhibo Zhou}received his B.Sc. degree in Communication Engineering from the School of Electronic Information and Communications, Huazhong University of Science and Technology, Wuhan, China, in 2021. He is currently pursuing his M.Sc. degree in Information and Communication Engineering with the School of Electronic Information and Communications, Huazhong University of Science and Technology, Wuhan, China. His research interests include analysis and applications of reconfigurable intelligent surface.
\end{IEEEbiography}
\begin{IEEEbiography}
    [{\includegraphics[width=1in,height=1.25in,clip,keepaspectratio]{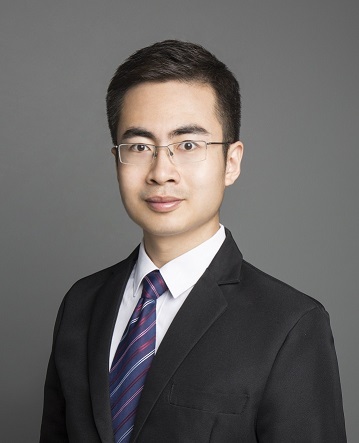}}]{Haifan Yin} (Senior Member, IEEE) received the B.Sc. degree in electrical and electronic engineering and the M.Sc. degree in electronics and information engineering from the Huazhong University of Science and Technology, Wuhan, China, in 2009 and 2012, respectively, and the Ph.D. degree from Télécom ParisTech in 2015. From 2009 to 2011, he was a Research and Development Engineer with the Wuhan National Laboratory for Optoelectronics, Wuhan, working on the implementation of TD-LTE systems. From 2016 to 2017, he was a DSP Engineer at Sequans Communications (IoT chipmaker), Paris, France. From 2017 to 2019, he was a Senior Research Engineer working on 5G standardization at Shanghai Huawei Technologies Company Ltd., where he has made substantial contributions to 5G standards, particularly the 5G codebooks. Since May 2019, he has been a Full Professor with the School of Electronic Information and Communications, Huazhong University of Science and Technology. His current research interests include 5G and 6G networks, signal processing, machine learning, and massive MIMO systems. He was the National Champion of 2021 High Potential Innovation Prize awarded by the Chinese Academy of Engineering, a recipient of the China Youth May Fourth Medal (the top honor for young Chinese), and a recipient of the Stephen O. Rice Prize.
\end{IEEEbiography}
\begin{IEEEbiography}                               [{\includegraphics[width=1in,height=1.25in,clip,keepaspectratio]{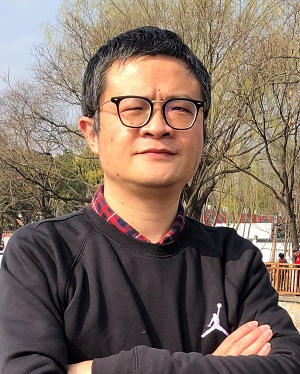}}]{Li Tan} received the B.Sc. degree in Telecommunications Engineering, the M.Sc. degree in Electrical Circuit and System, and the D.Sc. degree in Information and Communication Engineering from Huazhong University of Science and Technology, Wuhan, China, in 1999, 2002, and 2009, respectively. Since July 1999, he has been a Lecturer with Huazhong University of Science and Technology. His current research interests include wireless communication, radio resource management, energy efficiency, reconfigurable intelligent surface, stochastic computing, and probabilistic CMOS.
\end{IEEEbiography}
\begin{IEEEbiography}
[{\includegraphics[width=1in,height=1.25in,clip,keepaspectratio]{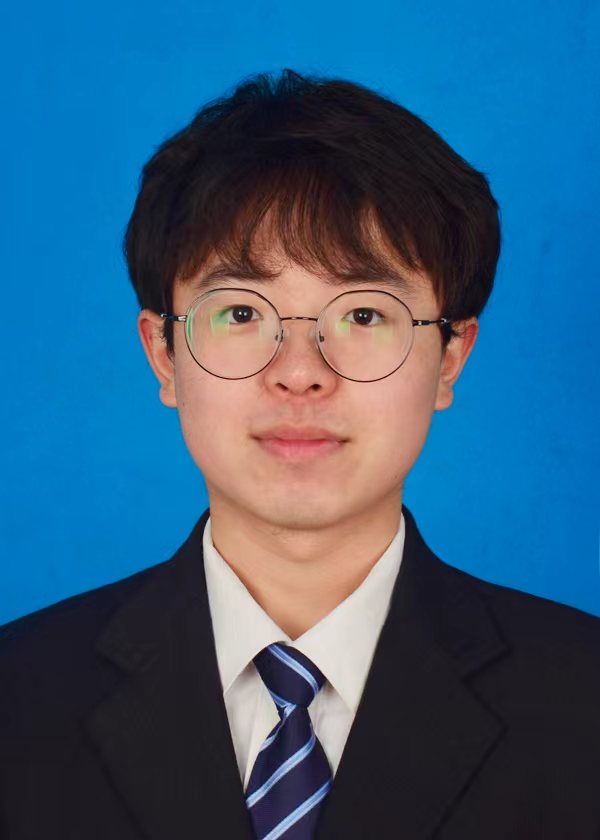}}]{Ruikun Zhang} received the B.Sc. degree in electronic science and technology from the School of Electronic Information and Communications, Huazhong University of Science and Technology, Wuhan, China, in 2023, where he is currently a graduate student with the School of Electronic Information and Communications. His research interests include analysis and applications of RIS and MIMO.
\end{IEEEbiography}

\begin{IEEEbiography}
    [{\includegraphics[width=1in,height=1.25in,clip,keepaspectratio]{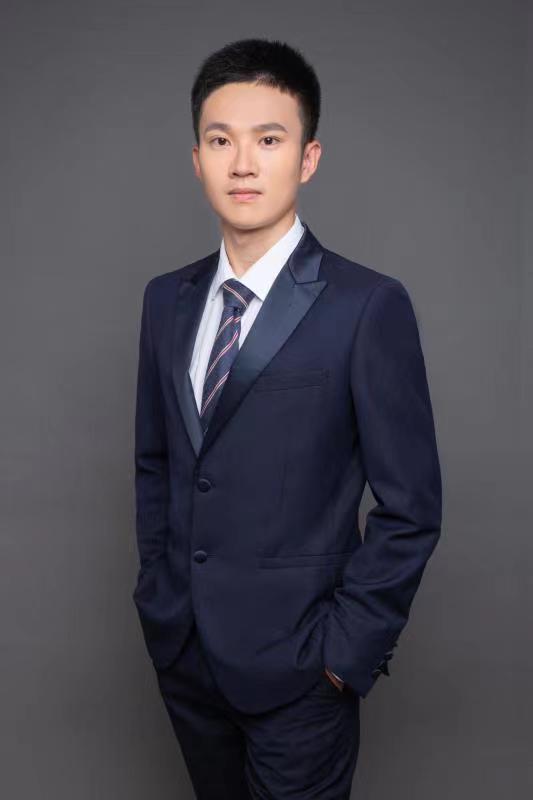}}] {Kai Wang} received the B.Sc. degree in electronic and information engineering from Huazhong University of Science and Technology, Wuhan, China, in 2021, where he is currently pursuing the M.Sc. degree in information and communications engineering. His current research interests include communication systems and reconfigurable intelligent surface.
\end{IEEEbiography}

\begin{IEEEbiography}
    [{\includegraphics[width=1in,height=1.25in,clip,keepaspectratio]{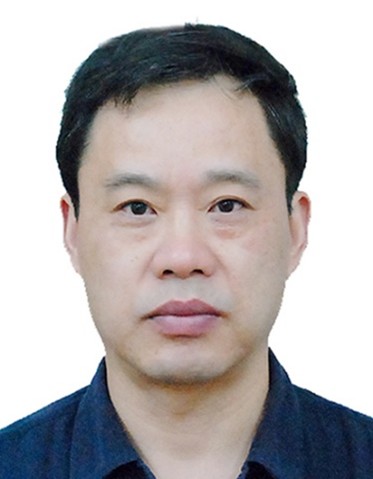}}] {Yingzhuang Liu} received his Ph.D. degree from Huazhong University of Science and Technology in 2000, and completed his postdoctoral research at Université Paris-Sud 11 in December 2001. He currently serves as a professor at the School of Electronic Information and Communications, Huazhong University of Science and Technology, where he is also the director of the Telecommunications Department and the head of the Broadband Wireless Communication Team. Additionally, he serves as the director of the HUST-China Mobile 5G Joint Laboratory and is a member of the European Organization for Nuclear Research (CERN) Large Hadron Collider (LHC) ALICE International Collaboration Group, where he serves as the Chinese group technology coordinator. He is also a member of the IEEE 802.11ax International Standardization Group, He has long been engaged in wireless communication research and development, having led several significant national projects in this field, including key projects funded by the National Natural Science Foundation of China, major national science and technology projects, and international cooperation projects. He has published over 100 papers, applied for more than 60 patents (including nearly 30 international patents), and submitted and passed over 10 technical proposals to mobile communication standardization organizations such as IMT Advanced/Long-Term Evolution Plus (LTE+)/China Communications Standards Association (CCSA).
\end{IEEEbiography}

\end{document}